\documentclass{emulateapj}
\usepackage{amssymb}
\usepackage{apjfonts}
\usepackage{graphicx}
\usepackage{natbib}
\usepackage{multirow}
\newcommand{\lya}{Ly~$\alpha$}
\def\SiIV{Si\,{\sevenrm\,IV}}

\def\CIV{C\,{\sc iv}}
\newcommand{\CIVa}{C{\sevenrm~IV}\,$\lambda$1548}
\newcommand{\CIVb}{C{\sevenrm~IV}\,$\lambda$1551}

\newcommand{\CIVwave}{C{\sevenrm~IV}\,$\lambda$1549}

\def\MgII{Mg\,{\sc ii}}

\newcommand{\OIIwave}{[O{\sevenrm\, II}]\,$\lambda$3728}
\newcommand{\OIwave}{O{\sevenrm\, I}\,$\lambda$1302}
\newcommand{\SiIIwave}{Si{\sevenrm\, II}\,$\lambda$1304}

\def\NV{N\,{\sc v}}
\def\OVI{O\,{\sc vi}}

\def\zabs{$z_{\rm abs}$}
\def\zem{$z_{\rm em}$}
\def\kms{$\rm km\,s^{-1}$}

 \font\sevenrm=cmr7 scaled 1000

\begin{document}
\title{The correlated variations of \CIV\ narrow absorption lines and quasar continuum}
\shorttitle{The correlated variations of absorption lines and quasar continuum}
\shortauthors{Chen et al.}

\author{Zhi-Fu Chen\altaffilmark{1,2}, Ting-Ting Pang\altaffilmark{2}, Bing He\altaffilmark{2}, and Yong Huang\altaffilmark{2}}
\altaffiltext{1}{College of Science, Guangxi University for Nationalities, Nanning 530006, China}
\altaffiltext{2}{School of Materials Science and Engineering, Baise University, Baise 533000, China; zhichenfu@126.com}

\begin{abstract}
We assemble 207 variable quasars from the Sloan Digital Sky Survey, all with at least 3 observations, to analyze \CIV\ narrow absorption doublets, and obtain 328 \CIV\ narrow absorption line systems. We find that 19 out of 328 \CIV\ narrow absorption line systems were changed by $\mid\Delta W_r^{\lambda1548}\mid\ge3\sigma_{\rm \Delta W_r^{\lambda1548}}$ on timescales from 15.9 to 1477 days at rest-frame. Among the 19 obviously variable \CIV\ systems, we find that (1) 14 systems have relative velocities $\upsilon_r>0.01c$ and 4 systems have $\upsilon_r>0.1c$, where $c$ is the speed of light; (2) 13 systems are accompanied by other variable \CIV\ systems; (3) 9 systems were changed continuously during multiple observations; and (4) 1 system with $\upsilon_{\rm r}=16~862$ \kms\ was enhanced by $\Delta W_r^{\lambda1548}=2.7\sigma_{\rm \Delta W_r^{\lambda1548}}$ in 0.67 day at rest-frame. The variations of absorption lines are inversely correlated with the changes in the ionizing continuum. We also find that large variations of \CIV\ narrow absorption lines are difficult to form on a short timescale.
\end{abstract}

\section{Introduction}
It is now widely accepted that feedback is a very important mechanism during the formation and evolution of galaxies. Outflow is a foundational component of active galactic nuclei, and is related to the gas lifted off the central accretion disk. In addition to regulating the growth of supermassive black holes, through feedback the outflow affects the multiphase distribution of the surrounding gas, influences the chemical and dynamical evolution of the galaxy, regulates the enrichment of the intergalactic medium, and enhances/quenches star formation within the host galaxy.

It is found that many AGN spectra exhibit asymmetric or even double-peaked features in the emission-line profiles \cite[e.g.,][]{2010ApJ...708..427L,2013ApJ...769...95B,2014RAA....14.1234S,2016MNRAS.463...24L,2017NatAs...1..727K}, which can be ascribed to (1) outflows \cite[e.g.,][]{2005MNRAS.356.1029B,2008ApJ...680..926K}; (2) binary supermassive black holes (SMBHs) \cite[e.g.,][]{2009Natur.458...53B,2010Natur.463E...1G,2010ApJ...725..249S}; and (3) accretion disk emission lines \cite[e.g.,][]{1989ApJ...339..742C,2006ApJ...652..112C,2009ApJ...695.1227L}. The emission lines with asymmetric profiles are a powerful and fashionable tool to investigate the properties of outflows. We also note that quasar absorption lines with $z_{\rm abs}<z_{\rm em}$ are very common, which could be originated in (1) foreground galaxies that are beyond the gravitational well of the quasar system \cite[e.g.,][]{1986A&A...155L...8B,1991A&A...243..344B,2006ApJ...645L.105B,2014MNRAS.441..886F,2016MNRAS.457..267L}; (2) quasar outflows \cite[e.g.,][]{2011MNRAS.410.1957H,2013MNRAS.435..133H,2015ApJ...799...63C,2016MNRAS.462.3285P}; (3) quasar surrounding environments, such as quasar's host galaxy, circumgalactic medium (CGM), and galaxy cluster. The foreground galaxies generally produce absorption lines with $z_{\rm abs} \ll z_{\rm em}$. The quasar surrounding environments often make absorption lines with $z_{\rm abs} \approx z_{\rm em}$. While, due to the relative motion of the clumpy clouds ejected by the quasar, absorption lines formed in quasar outflows generally show an absorption redshift $z_{\rm abs} \lesssim z_{\rm em}$. The blueshifted absorption lines are common in the quasar spectra and their detections don't depend on quasar emissions, so they are a very popular tool to study the outflow as well.

Broad absorption lines (BALs), which have line widths larger than a few thousand \kms\ at depth $>10\%$ below the continuum \cite[e.g.,][]{1979ApJ...234...33W}, are undoubtedly related to the quasars and likely formed in quasar outflows. While narrow absorption lines (NALs), which have line widths less than a few hundred \kms, can be originated in a wide variety of media. No matter what the relationship between the medium producing NALs and the quasar is, NALs often exhibit similar profiles of lines, especially in the low- or middle-resolution spectra. Therefore, it is not easy to distinguish outflow NALs from the NALs originated in other media in a single-epoch spectra. It has long been known that the quasar NALs can vary in equivalent width on timescales from months to years \cite[e.g.,][]{2004ApJ...613..129W,2004ApJ...601..715N,2011MNRAS.410.1957H,2013ApJ...777...56C,2014ApJ...792...77M,2015A&A...581A.109B,2015MNRAS.450.3904C,2016ApJ...825...25M}. The variations of NALs can be interpreted by (1) the change in the background ionizing continuum; (2) the change in the shielding gas located between the continuum source and absorbing clouds; (3) clumpy clouds moving across the continuum source. The line variation would be a good characteristic to manifest whether the NALs are formed in the outflow or not, since, based on above scenarios, it is unreality to expect obvious variation in the intervening and environmental NALs on a timescale of years.

While the origin of the variation in NALs is still an open question, the variations of absorption lines are a useful tool to constrain the physical and geometrical conditions, and/or kinematics of the absorbing clouds. In the photo-ionization scheme, the line variation is related to the flux density of the incident ionizing continuum, thus the timescale of the variation can set a limit on the recombination or ionization timescale.
In the scheme of the absorbing cloud moving across our sightline, the timescale of the variation can help to limit the transverse velocity of clumpy cloud across the ionizing continuum source. In this paper, using the quasars with multi-epoch spectra of the Sloan Digital Sky survey \cite[SDSS;][]{2000AJ....120.1579Y}, we will examine the variations of \CIV\ NALs, and investigate the relationship between the variations of the NALs and continuum. The \MgII\ and \SiIV\ NALs, and the BALs will be studied in our future work.

In Section \ref{sect:qsosample_spectra_analysis}, we describe the data sample and spectral analysis. We present the properties of variable \CIV\ systems and discussions in Section \ref{sect:discussion}. A summary is presented in Section \ref{sect:summary}. In this paper, we adopt the $\rm \Lambda CDM$ cosmology with $\rm \Omega_M=0.3$, $\rm \Omega_\Lambda=0.7$, and $\rm H_0=70~km~s^{-1}~Mpc^{-1}$.

\section{Quasar sample and spectral analysis}
\label{sect:qsosample_spectra_analysis}
\subsection{Quasar sample}
\label{sect:Quasar_sample}
 Through the ancillary science programs of the Baryon Oscillation Spectroscopic Survey \cite[BOSS;][]{2013AJ....145...10D,2016AJ....151...44D} and the Extended Baryon Oscillation Spectroscopic Survey \cite[eBOSS;][]{2016AJ....151...44D}, some quasars are repeatedly observed by the SDSS. The data release 14 quasar catalog from the eBOSS \cite[DR14Q; ][]{2017arXiv171205029P} is the latest SDSS catalog of the spectroscopically-confirmed quasars, which also includes previously spectroscopically-confirmed quasars from the SDSS. We select quasars from the DR14Q catalog, which have three- to six-epoch observations. Here, we don't consider the quasars with more than six observations, since they generally have a very short observation time intervals. The purpose of this paper is to analyse the variation of \CIV\ NALs, so we consider in our sample only quasars with $z_{\rm em}> 1.32$. Then, we download the quasar spectra from \url{https://data.sdss.org/sas/dr14/}. We aim to investigate the relationship between the variations of quasar radiations and absorption lines, thus we further limit the quasar sample with $\Delta f_{\rm 1350}\ge4\sigma_{\rm f_{1350}}$, where the $f_{\rm 1350}$ is the spectral flux at rest frame 1350 \AA. After these screenings, we obtain a sample of 207 quasars, which includes 675 spectra. For these 207 quasars, we determine the emission redshifts in the following sequence: (1) adopt the redshifts from our measurements of \OIIwave\ narrow emission lines when available; (2) use the improved redshifts of the SDSS quasars from \cite{2010MNRAS.405.2302H}; (3) adopt the \MgII\ emission line based redshifts from \cite{2017A&A...597A..79P} when available; and (4) or otherwise utilize the visual inspection redshifts of \cite{2017A&A...597A..79P}.

\subsection{Spectral analysis and parameter measurements}
\label{sect:Spectral_analysis}
In order to identify \CIV\ NALs, we adopt methods consistent with our previous works \cite[e.g.,][]{2014ApJS..210....7C,2014ApJS..215...12C,2015MNRAS.450.3904C} to analyze the quasar spectra, which are briefly described as follows.
\begin{enumerate}
  \item For each spectrum, we model a pseudo-continuum with a combination of cubic spline and multi-Gaussian functions in an iterative fashion, which is utilized to normalize the spectral flux and flux uncertainty. The NALs generally have line widths less than a few hundred \kms. Therefore, we first mark the continual absorption features with line widths larger than 1200 \kms\ at depths larger than 20\% under the pseudo-continuum fit \cite[e.g.;][]{2015MNRAS.450.3904C}. In the same spectral region, if none spectrum of a quasar shows an absorption feature less than 1200 \kms\ at depths larger than 20\% under the pseudo-continuum fit, we get rid of the marked absorption features. This step was carried out to account for NALs that might be evolved into mini-BALs or BALs.

  \item The pairs of \SiIIwave\ and \OIwave\ NALs at higher redshifts often lead to misidentifications of \CIV\ NALs at lower redshifts. In order to reduce the confusions arising from the \lya, \OIwave\ and \SiIIwave\ absorption features, we consider the spectra data, which are used to search for \CIV\ NALs, from 1310 \AA\ at rest frame to the red wing of the \CIVwave\ emission lines. The \CIV\ NALs are identified in the normalized spectra data, which are the spectral fluxes divided by the pseudo-continuum fit. A pair of Gaussian functions is invoked to model a \CIV\ doublet and the best fit is carried out with $\chi^2$ minimization. The model results are visually inspected one-by-one.
  \item The redshifts (\zabs) of \CIV\ NALs are yielded by the Gaussian function fitting centers of the \CIVa\ lines, and the equivalent widths ($W_{\rm r}$) of absorption lines at rest frame are determined by the integrations of the Gaussian function fitting profiles. The error ($\sigma_{\rm w}$) of the $W_{\rm r}$, which is contributed from the spectra flux uncertainty, is estimated by
      \begin{equation}
      \label{eq:werr}
      (1+z_{abs})\sigma_w=\frac{\sqrt{\sum\limits_{i=1}^{N}
      P^2(\lambda_i-\lambda_0)\frac{\sigma^2_{flux,i}}{f^2_{flux,i}}}}{\sum\limits_{i=1}^{N}
      P^2(\lambda_i-\lambda_0)}\Delta\lambda,
      \end{equation}
      where $P(\lambda_i-\lambda_0)$ is the absorption line profile centered at $\lambda_0$, $\lambda_i$ is the wavelength, $f_{flux,i}$ is the spectra flux, $\sigma_{flux,i}$ is the spectra flux uncertainty, and $N$ is the pixel number over $\rm \pm3\sigma$, where $\rm \sigma$ is determined by the fitting result of the Gaussian function. Due to the evolution of absorption lines or the signal-to-noise ratio of the spectra, we find that some \CIV\ NALs are remarkable in the spectra obtained at some times, but they are inconspicuous absorption features in the spectra obtained on the other times. For these inconspicuous absorption lines, we estimate their $\sigma_w$ within 200 \kms; these estimates are also considered as upper limits of absorption strengths.
\end{enumerate}

\begin{figure}
\centering
\includegraphics[width=0.42\textwidth]{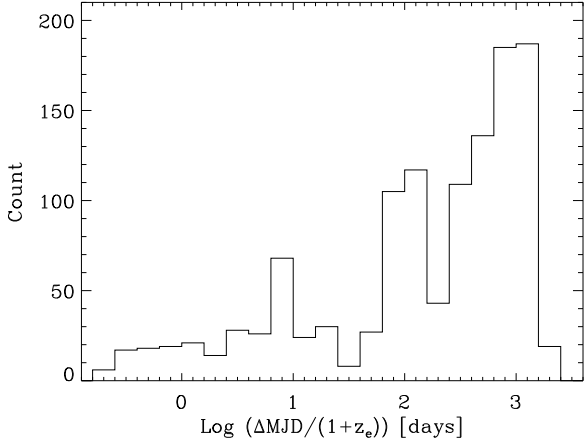}
\caption{Distribution of observed timescales of the 1210 \CIV\ NAL pairs. About 66\% of the timescales at rest-frame are located within the region of $\rm \Delta MJD >100$ days.}
\label{fig:delta_mjd}
\end{figure}

Among the 207 quasars, there are 166 quasars that at least one \CIV\ NAL system with $W_{\rm r}^{\lambda1548}\ge3\sigma_{\rm W_{\rm r}^{\lambda1548}}$ and $W_{\rm r}^{\lambda1551}\ge2\sigma_{\rm W_{\rm r}^{\lambda1551}}$ was detected in one of the multi-epoch spectra. We obtain 328 \CIV\ NAL systems, which constitute of 1210 \CIV\ NAL pairs. Here, a \CIV\ NAL pair is defined as the same one \CIV\ NAL system that is observed in two-epoch spectra. The observed timescales of the 1210 \CIV\ NAL pairs are shown in Figure \ref{fig:delta_mjd}, and their parameters are listed in Table \ref{Tab:abs}.

\begin{table*}[htbp]
\caption{Catalog of \CIV\ narrow absorption line systems} \tabcolsep 1.1mm \centering 
\label{Tab:abs}
 \begin{tabular}{ccccccccccccccc}
 \hline\hline\noalign{\smallskip}
ID&SDSS NAME & PLATE & MJD & FIBER & \zem & \zabs & $W_{\rm  r}^{\lambda1548}$ & $W_{\rm r}^{\lambda1551}$ & $f_{\rm 1350}$&$v_{\rm r}$&ID&symbol&SL\\
&&&&&&&\AA&\AA& $10^{-17}~erg~s^{-1}~cm^{-2}$ \AA$^{-1}$ &\kms\\
(1)&(2)&(3)&(4)&(5)&(6)&(7)&(8)&(9)&(10)&(11)&(12)&(13)&(14)\\
\hline\noalign{\smallskip}
1	&	001142.72+255537.1	&	6880	&	56543	&	865	&	2.2944 	&	2.2466	&	0.43 	$\pm$	0.09 	&	0.34 	$\pm$	0.09 	&	3.805 	$\pm$	0.523 	&	4384	&	1	&	0	&	0.4	\\
1	&	001142.72+255537.1	&	6276	&	56269	&	370	&	2.2944 	&	2.2462	&	0.38 	$\pm$	0.08 	&	0.47 	$\pm$	0.10 	&	5.124 	$\pm$	0.622 	&	4421	&	1	&		&		\\
2	&	001142.72+255537.1	&	7664	&	57367	&	859	&	2.2944 	&	2.2467	&	1.15 	$\pm$	0.33 	&	0.35 	$\pm$	0.15 	&	1.027 	$\pm$	0.383 	&	4375	&	1	&	0	&	2.3	\\
2	&	001142.72+255537.1	&	6276	&	56269	&	370	&	2.2944 	&	2.2462	&	0.38 	$\pm$	0.08 	&	0.47 	$\pm$	0.10 	&	5.124 	$\pm$	0.622 	&	4421	&	1	&		&		\\
3	&	001142.72+255537.1	&	7664	&	57367	&	859	&	2.2944 	&	2.2467	&	1.15 	$\pm$	0.33 	&	0.35 	$\pm$	0.15 	&	1.027 	$\pm$	0.383 	&	4375	&	1	&	0	&	2.1	\\
3	&	001142.72+255537.1	&	6880	&	56543	&	865	&	2.2944 	&	2.2466	&	0.43 	$\pm$	0.09 	&	0.34 	$\pm$	0.09 	&	3.805 	$\pm$	0.523 	&	4384	&	1	&		&		\\
4	&	001142.72+255537.1	&	6880	&	56543	&	865	&	2.2944 	&	2.2077	&	0.70 	$\pm$	0.14 	&	0.66 	$\pm$	0.20 	&	3.805 	$\pm$	0.523 	&	7999	&	2	&	0	&	0.6	\\
4	&	001142.72+255537.1	&	6276	&	56269	&	370	&	2.2944 	&	2.2075	&	0.58 	$\pm$	0.16 	&	0.57 	$\pm$	0.18 	&	5.124 	$\pm$	0.622 	&	8017	&	2	&		&		\\
5	&	001142.72+255537.1	&	7664	&	57367	&	859	&	2.2944 	&	2.2074	&	0.64 	$\pm$	0.64 	&	0.64 	$\pm$	0.64 	&	1.027 	$\pm$	0.383 	&	8027	&	2	&	0	&	0.1	\\
5	&	001142.72+255537.1	&	6880	&	56543	&	865	&	2.2944 	&	2.2077	&	0.70 	$\pm$	0.14 	&	0.66 	$\pm$	0.20 	&	3.805 	$\pm$	0.523 	&	7999	&	2	&		&		\\
6	&	001142.72+255537.1	&	7664	&	57367	&	859	&	2.2944 	&	2.2074	&	0.64 	$\pm$	0.64 	&	0.64 	$\pm$	0.64 	&	1.027 	$\pm$	0.383 	&	8027	&	2	&	0	&	0.1	\\
6	&	001142.72+255537.1	&	6276	&	56269	&	370	&	2.2944 	&	2.2075	&	0.58 	$\pm$	0.16 	&	0.57 	$\pm$	0.18 	&	5.124 	$\pm$	0.622 	&	8017	&	2	&		&		\\
7	&	001142.72+255537.1	&	7664	&	57367	&	859	&	2.2944 	&	2.2823	&	1.14 	$\pm$	0.27 	&	1.26 	$\pm$	0.27 	&	1.027 	$\pm$	0.383 	&	1103	&	3	&	0	&	0.4	\\
7	&	001142.72+255537.1	&	6880	&	56543	&	865	&	2.2944 	&	2.2825	&	1.04 	$\pm$	0.09 	&	1.04 	$\pm$	0.09 	&	3.805 	$\pm$	0.523 	&	1085	&	3	&		&		\\
8	&	001142.72+255537.1	&	6880	&	56543	&	865	&	2.2944 	&	2.2825	&	1.04 	$\pm$	0.09 	&	1.04 	$\pm$	0.09 	&	3.805 	$\pm$	0.523 	&	1085	&	3	&	0	&	1.8	\\
8	&	001142.72+255537.1	&	6276	&	56269	&	370	&	2.2944 	&	2.2823	&	1.27 	$\pm$	0.09 	&	1.00 	$\pm$	0.08 	&	5.124 	$\pm$	0.622 	&	1103	&	3	&		&		\\
9	&	001142.72+255537.1	&	7664	&	57367	&	859	&	2.2944 	&	2.2823	&	1.14 	$\pm$	0.27 	&	1.26 	$\pm$	0.27 	&	1.027 	$\pm$	0.383 	&	1103	&	3	&	0	&	0.5	\\
9	&	001142.72+255537.1	&	6276	&	56269	&	370	&	2.2944 	&	2.2823	&	1.27 	$\pm$	0.09 	&	1.00 	$\pm$	0.08 	&	5.124 	$\pm$	0.622 	&	1103	&	3	&		&		\\
\hline\hline\noalign{\smallskip}
\end{tabular}
\begin{flushleft}
\footnote[]~Column (1): identification number of a \CIV\ NAL pairs. Column (2): SDSS object name. Column (3)---(5): PlateID, MJD, and FiberID of the SDSS spectra. Column (6): emission-line redshift of the quasar. Column (7): absorption line redshift of the \CIV\ NAL system. Column (8): rest-frame equivalent width and corresponding error of the \CIVa. Column (9): rest frame equivalent width and corresponding error of the \CIVb. Column (10): the 1350 \AA\ continuum flux and flux uncertainty measured from the original spectra of the SDSS. Column (11): velocity offset of the \CIV\ NAL system with respect to the quasar emission line redshift. Column (12): identification number of the \CIV\ NAL system. Column (13): 1 indicates enhanced \CIV\ NALs with $\mid\Delta W_r^{\lambda1548}\mid\ge3\sigma_{\rm \Delta W_r^{\lambda1548}}$; -1 indicates weakened \CIV\ NALs with $\mid\Delta W_r^{\lambda1548}\mid\ge3\sigma_{\rm \Delta W_r^{\lambda1548}}$; 2 indicates emerged \CIV\ NALs; -2 indicates disappeared \CIV\ NALs; and 0 indicates \CIV\ NALs with $\mid\Delta W_r^{\lambda1548}\mid<3\sigma_{\rm \Delta W_r^{\lambda1548}}$. See Equations \ref{eq:delta_w} and \ref{eq:delta_w_err} for the calculations of the $\Delta W_{\rm r}^{\lambda1548}$ and $\sigma_{\Delta W_r^{\lambda1548}}$. Column (14): the significance level (SL) of the variation, which is equated to $\mid\Delta W_r^{\lambda1548}\mid/\sigma_{\rm \Delta W_r^{\lambda1548}}$. Note that, if the rest frame equivalent widths and corresponding errors of both lines of a \CIV\ doublet are same, these values are the 1 $\sigma$ upper limits of absorption strengths and indicate that the \CIV\ NALs are not observed in the corresponding spectra. This table is available in its entirety in machine-readable form.
\end{flushleft}
\end{table*}

\section{The properties of variable \CIV\ systems and discussions}
\label{sect:discussion}
\subsection{The variations of \CIV\ NALs and UV continuum}
\label{sect:variation}
\begin{figure*}
\centering
\includegraphics[width=0.45\textwidth]{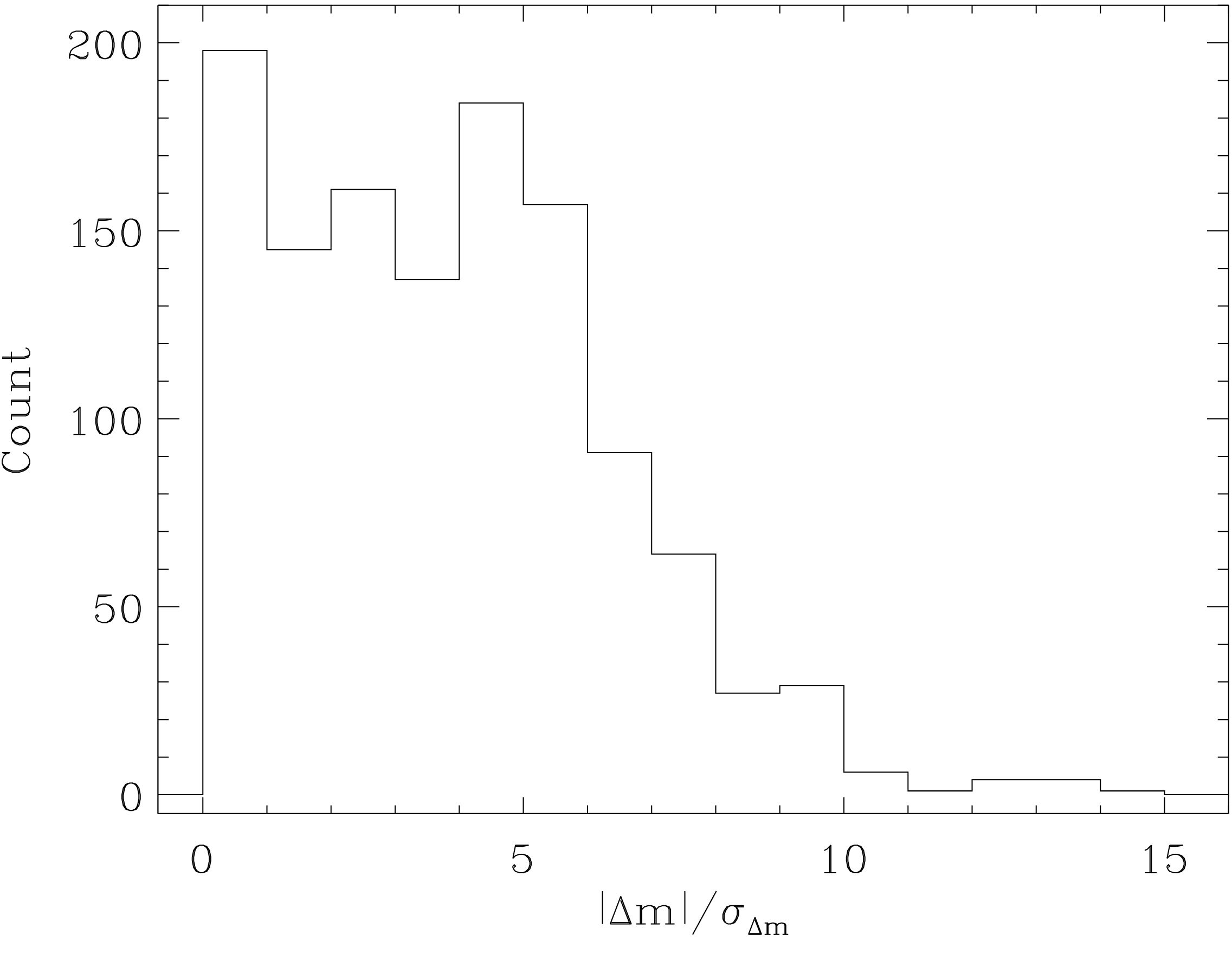}
\hspace{4ex}
\includegraphics[width=0.45\textwidth]{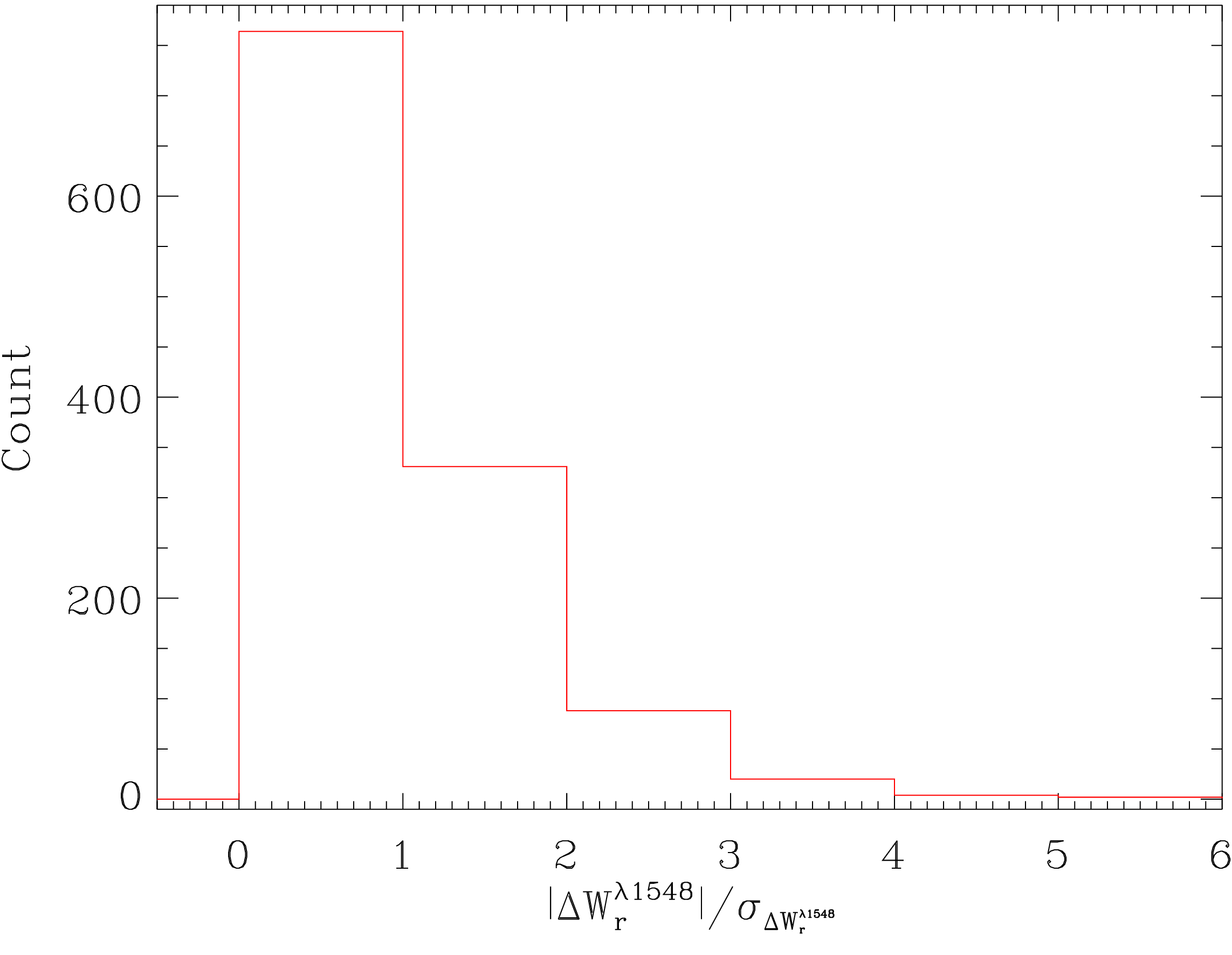}
\caption{The normalized differences of the 1350 \AA\ continuum fluxes (left panel) and the absorption strengths $W_{\rm r}^{\lambda1548}$ (right panel) at two epochs. About $58\%$ of the \CIV\ NAL-pairs are fallen into the region of $\mid\Delta m\mid\ge3\sigma_{\rm \Delta m}$, and there are 27 \CIV\ NAL-pairs with $\mid\Delta W_r^{\lambda1548}\mid\ge3\sigma_{\rm \Delta W_r^{\lambda1548}}$.}
\label{fig:delta_w_f}
\end{figure*}

Section \ref{sect:Quasar_sample} claims that our data sample is limited to the quasars with $f_{\rm 1350}\ge4\sigma_{\rm f_{1350}}$, which is the most significant variation among the multiple-spectra pairs of a quasar. We note that some authors might fit a power-law continuum for each spectrum and derive the continuum flux at a band by directly measuring the fitting power-law continuum, which is unnecessary for this paper. We can derive the continuum flux at 1350 \AA\ ($f_{\rm 1350}$) by directly measuring the SDSS original spectra. We believe that the SDSS original spectra and a fitting power-law continuum would not produce obviously different $f_{\rm 1350}$, especially for the differences between the $f_{\rm 1350}$ measured from the two-epoch spectra of a quasar. Now we start to compute the flux variations of the 1350 \AA\ continuum for our 1210 \CIV\ NAL pairs. Here, we converse the fluxes to magnitudes, and define the basic variation as
\begin{equation}
\label{eq:delta_mag}
\Delta m = -2.5\times log(\frac{f_{1350}^1}{f_{1350}^2}),
\end{equation}
and the corresponding error as
\begin{equation}
\label{eq:delta_mag_err}
\sigma_{\Delta m} =\sqrt{(\frac{\sigma_{f_{1350}^1}}{f_{1350}^1})^2 + (\frac{\sigma_{f_{1350}^2}}{f_{1350}^2})^2} ,
\end{equation}
where the $f_{\rm 1350}^1$, $f_{\rm 1350}^2$ are the observed flux of the 1350 \AA\ continuum at two epochs, and the $\sigma_{\rm f_{1350}^1}$, $\sigma_{\rm f_{1350}^2}$ are the corresponding errors. The normalized differences of the 1350 \AA\ continuum fluxes are shown in the left panel of Figure \ref{fig:delta_w_f}. We find that a significant variation of $\mid\Delta m\mid\ge3\sigma_{\rm \Delta m}$ is happened to about $58\%$ of the 1210 \CIV\ NAL-pairs. We note that the spectrophotometric calibration of the SDSS has an error of about 5\% at r-band \cite[e.g.,][]{2008ApJS..175..297A,2016ApJ...831..157M,2016AJ....151....8Y,2017arXiv170709322A}. Considering 5\% of spectral flux as the error of flux calibration, we also find that 49\% of the 1210 \CIV\ NAL-pairs show variations with $\mid\Delta m\mid\ge3\sigma_{\rm \Delta m}$. This paper aims to investigate the relationship between the variations of absorption lines and quasar continuum, which is expected to be diminished rather than enhanced by the flux calibration uncertainty. Therefore, we don't consider the flux calibration uncertainty throughout this paper.

To investigate the variations of \CIV\ NALs, we compute the differences of absorption strengths via
\begin{equation}
\label{eq:delta_w}
\Delta W_r^{\lambda} = W_{r,1}^{\lambda} - W_{r,2}^{\lambda},
\end{equation}
and corresponding errors via
\begin{equation}
\label{eq:delta_w_err}
\sigma_{\Delta W_r^{\lambda}} = \sqrt{ \sigma_{W_{r,1}^{\lambda}}^2 + \sigma_{W_{r,2}^{\lambda}}^2 } ,
\end{equation}
where the $W_{r,1}^{\lambda}$, $W_{r,2}^{\lambda}$ are the equivalent widths at rest frame at two epochs, and the $\sigma_{W_{r,1}^{\lambda}}$, $\sigma_{W_{r,2}^{\lambda}}$ are corresponding errors. The normalized differences of the absorption strengths are shown in the right panels of Figure \ref{fig:delta_w_f}. We find that there are 19 \CIV\ NAL systems with $\mid\Delta W_r^{\lambda1548}\mid\ge3\sigma_{\rm \Delta W_r^{\lambda1548}}$, which are imprinted in the spectra of 16 quasars. These 19 \CIV\ NAL systems with significant variations constitute of 72 \CIV\ NAL-pairs, of which 27 pairs have variations of $\mid\Delta W_r^{\lambda1548}\mid\ge3\sigma_{\rm \Delta W_r^{\lambda1548}}$.

\cite{2004ApJ...613..129W} had investigated the variations of associated \CIV, \NV, and \OVI\ NALs with $\upsilon_{\rm r}\le 5000$ \kms, and found that 4 out of 15 quasars, or 4 out of 19 NALs, contained variable NALs, which indicates that about 21\% associated NALs are variable. With a limit of $\upsilon_{\rm r}\le 5000$ \kms, our sample contains 7 out of 80 quasars, or 7 out of 100 NALs, and include variable \CIV\ NALs, which means that about 7\% \CIV\ NALs are variable. The fraction of variable NALs is much less than that reported by \cite{2004ApJ...613..129W}, which might be mainly ascribed to the reason that \cite{2004ApJ...613..129W} contained \CIV, \NV, and \OVI\ NALs, while we only consider \CIV\ NALs.

\subsection{The variable \CIV\ narrow absorption line systems}
\label{sect:variable_NALs}
In this section, we describe the 19 obviously variable \CIV\ NALs with $\mid\Delta W_r^{\lambda1548}\mid\ge3\sigma_{\rm \Delta W_r^{\lambda1548}}$ one-by-one, whose spectra are shown in Figure \ref{fig:variableabs}.
\begin{enumerate}
  \item Quasar SDSS J003135.57+003421.2 with $z_{\rm em}=2.2300$. It was spectroscopically observed by the SDSS on $\rm MJD~=~52262$, 55182, 55443 and 57006. In the quasar spectra, we detect 4 \CIV\ NALs with $z_{\rm abs}=1.7333$, 1.9979, 2.0082 and 2.0246, and $\upsilon_{\rm r}=49~588$, $22~329$, $21~306$ and $19~682$ \kms\ from the quasar system. A weakened system with $\Delta W_r^{\lambda1548}=-3.9\sigma_{\rm \Delta W_r^{\lambda1548}}$ had $\upsilon_{\rm r}=21~306$ \kms. The system with $\upsilon_{\rm r}=22~329$ \kms\ also showed a variation of $\Delta W_r^{\lambda1548}=-2.8\sigma_{\rm \Delta W_r^{\lambda1548}}$. Note that the negative value of the significance level indicates a weakened system, and a vice versa for the oppositive value. These two variable \CIV\ systems were weakened on the same timescale.

  \item Quasar SDSS J004323.43-001552.4 with $z_{\rm em}=2.8200$. It was spectroscopically observed by the SDSS on $\rm MJD~=~55184$, 55186, 55444 and 57016. In the quasar spectra, we detect 5 \CIV\ NALs with $z_{\rm abs}=2.4501$, 2.7882, 2.7993, 2.8145 and 2.8342, and $\upsilon_{\rm r}=30~448$, $2507$, $1630$, $432$ and $-1113$ \kms\ from the quasar system. A system with $\upsilon_{\rm r}=2507$ \kms\ was enhanced first by $\Delta W_r^{\lambda1548}=2.6\sigma_{\rm \Delta W_r^{\lambda1548}}$ from $\rm MJD=55184$ to 55444, and then was weakened by $\Delta W_r^{\lambda1548}=-3.4\sigma_{\rm \Delta W_r^{\lambda1548}}$ from $\rm MJD=5444$ to 57106. The other one system with $\upsilon_{\rm r}=-1113$ \kms\ also showed a variation of $\Delta W_r^{\lambda1548}=-2.2\sigma_{\rm \Delta W_r^{\lambda1548}}$ from $\rm MJD=55184$ to 57106, whose weakened timescale was longer than that of the system with $\upsilon_{\rm r}=2507$.

  \item Quasar SDSS J004856.34+005648.1 with $z_{\rm em}=2.3230$. It was spectroscopically observed by the SDSS on $\rm MJD~=~51913$, 55201 and 55451. In the quasar spectra, we detect 3 \CIV\ NALs with $z_{\rm abs}=1.8929$, 2.2651 and 2.3052, and $\upsilon_{\rm r}=43~318$, $5272$ and $1611$ \kms\ from the quasar system. The system with $\upsilon_{\rm r}=1611$ \kms\ was weakened first by $\Delta W_r^{\lambda1548}=-3.8\sigma_{\rm \Delta W_r^{\lambda1548}}$ from $\rm MJD=51913$ to 55201, and then was enhanced by $\Delta W_r^{\lambda1548}=1.6\sigma_{\rm \Delta W_r^{\lambda1548}}$ from $\rm MJD=55201$ to 55451.

  \item Quasar SDSS J015017.70+002902.4 with $z_{\rm em}=2.9980$. It was spectroscopically observed by the SDSS on $\rm MJD~=~51793$, 55182, 55447 and 56900. In the quasar spectra, we detect 4 \CIV\ NALs with $z_{\rm abs}=2.8121$, 2.8345, 3.0066 and 3.0189, and $\upsilon_{\rm r}=14~218$, $12~519$, $-644$ and $-1564$ \kms\ from the quasar system. The obviously variable system had $\upsilon_{\rm r}=14~218$ \kms, which was emerged first during $\rm MJD=51793$ and 55182, and then was weakened by $\Delta W_r^{\lambda1548}=-4.3\sigma_{\rm \Delta W_r^{\lambda1548}}$ from $\rm MJD=55182$ to 56900. The variations of the system with $\upsilon_{\rm r}=12~519$ \kms\ were consistent with the system with $\upsilon_{\rm r}=14~218$ \kms. Another one variable system with $\upsilon_{\rm r}=-644$ \kms\ was enhanced first by $\Delta W_r^{\lambda1548}=2\sigma_{\rm \Delta W_r^{\lambda1548}}$ from $\rm MJD=51793$ to 55447, and then was weakened by $\Delta W_r^{\lambda1548}=-2.5\sigma_{\rm \Delta W_r^{\lambda1548}}$ from $\rm MJD=55447$ to 56900.

  \item Quasar SDSS J024457.18-010809.8 with $z_{\rm em}=3.9780$. It was spectroscopically observed by the SDSS on $\rm MJD~=~51871$, 55247, 55455 and 57041. In the quasar spectra, we detect 3 \CIV\ NALs with $z_{\rm abs}=3.3209$, 3.6049 and 3.8078, and $\upsilon_{\rm r}=42~187$, $23~324$ and $10~432$ \kms\ from the quasar system. The significantly variable \CIV\ system with $\Delta W_r^{\lambda1548}=-3\sigma_{\rm \Delta W_r^{\lambda1548}}$ was located at $z_{\rm abs}=3.3209$ and had $\upsilon_{\rm r}=42~187$ \kms.

  \item Quasar SDSS J025911.85+001812.8 with $z_{\rm em}=2.8630$. It was spectroscopically observed by the SDSS on $\rm MJD~=~53742$, 55476 and 56984. In the quasar spectra, we detect 3 \CIV\ NALs with $z_{\rm abs}=2.7836$, 2.8225 and 2.8369, and $\upsilon_{\rm r}=6229$, $3161$ and $2033$ \kms\ from the quasar system. The system with $\upsilon_{\rm r}=2033$ \kms\ was enhanced by $\Delta W_r^{\lambda1548}=3.2\sigma_{\rm \Delta W_r^{\lambda1548}}$ from $\rm MJD=53742$ to 55476.

  \item Quasar SDSS J081435.19+502946.3 with $z_{\rm em}=3.8800$. It was spectroscopically observed by the SDSS on $\rm MJD~=~55180$, 55517 and 55590. In the quasar spectra, we detect 3 \CIV\ NALs with $z_{\rm abs}=3.2818$, 3.3361 and 3.5576, and $\upsilon_{\rm r}=39~009$, $35~286$ and $20~472$ \kms\ from the quasar system. The significantly variable system with $\Delta W_r^{\lambda1548}=-4\sigma_{\rm \Delta W_r^{\lambda1548}}$ and $\upsilon_{\rm r}=20~472$ \kms\ was continuously weaken during $\rm MJD=55180$ and 55590. The other one slightly weakened system had $\upsilon_{\rm r}=35~286$ \kms.

  \item Quasar SDSS J084525.84+072222.3 with $z_{\rm em}=2.3270$. It was spectroscopically observed by the SDSS on $\rm MJD~=~52964$, 55893 and 55946. In the quasar spectra, we detect 2 \CIV\ NALs with $z_{\rm abs}=2.0293$ and 2.3134, and $\upsilon_{\rm r}=28~039$ and $1228$ \kms\ from the quasar system. The system with $\upsilon_{\rm r}=1228$ \kms\ was enhanced first by $\Delta W_r^{\lambda1548}=2.9\sigma_{\rm \Delta W_r^{\lambda1548}}$ from $\rm MJD=52964$ to 55893, and then was weakened by $\Delta W_r^{\lambda1548}=-3.1\sigma_{\rm \Delta W_r^{\lambda1548}}$ during $\rm MJD=55893$ and 55964. The other one significantly variable system with $\upsilon_{\rm r}=28~039$ \kms\ was enhanced by $\Delta W_r^{\lambda1548}=2.4\sigma_{\rm \Delta W_r^{\lambda1548}}$ from $\rm MJD=52964$ to 55893.

  \item Quasar SDSS J115911.52+313427.2 with $z_{\rm em}=3.0380$. It was spectroscopically observed by the SDSS on $\rm MJD~=~53474$, 55589 and 56363. In the quasar spectra, we detect 3 \CIV\ NALs with $z_{\rm abs}=2.8624$, 2.9900 and 3.0272, and $\upsilon_{\rm r}=13~329$, $3587$ and $803$ \kms\ from the quasar system. The significantly variable system with $\Delta W_r^{\lambda1548}=3.5\sigma_{\rm \Delta W_r^{\lambda1548}}$ and $\upsilon_{\rm r}=3587$ \kms\ was continuously enhanced during $\rm MJD=53474$ and 56363.

  \item Quasar SDSS J120206.80+370919.5 with $z_{\rm em}=2.4782$. It was spectroscopically observed by the SDSS on $\rm MJD~=~53467$, 55621 and 57428. In the quasar spectra, we detect 2 \CIV\ NALs with $z_{\rm abs}=2.3444$ and 2.3737, and $\upsilon_{\rm r}=11~762$ and $9148$ \kms\ from the quasar system. The significantly variable system with $\Delta W_r^{\lambda1548}=3.5\sigma_{\rm \Delta W_r^{\lambda1548}}$ and $\upsilon_{\rm r}=11~762$ \kms\ was continuously enhanced during $\rm MJD=53467$ and 57428. The system with $\upsilon_{\rm r}=9148$ \kms\ was also slightly enhanced by $\Delta W_r^{\lambda1548}=1.7\sigma_{\rm \Delta W_r^{\lambda1548}}$ during the same timescale.

  \item Quasar SDSS J121347.74+373726.8 with $z_{\rm em}=1.7980$. It was spectroscopically observed by the SDSS on $\rm MJD~=~53472$, 55621 and 57426. In the quasar spectra, we detect one \CIV\ NALs with $z_{\rm abs}=1.7976$ and $\upsilon_{\rm r}=42$ \kms\ from the quasar system. This system was weakened first by $\Delta W_r^{\lambda1548}=-2.5\sigma_{\rm \Delta W_r^{\lambda1548}}$ during $\rm MJD=53472$ and 55621, and then obivously enhanced by $\Delta W_r^{\lambda1548}=3.8\sigma_{\rm \Delta W_r^{\lambda1548}}$ during $\rm MJD=55621$ and 57426.

  \item Quasar SDSS J121400.79+370936.7 with $z_{\rm em}=1.9233$. It was spectroscopically observed by the SDSS on $\rm MJD~=~53472$, 55591 and 57476. In the quasar spectra, we detect 2 \CIV\ NALs with $z_{\rm abs}=1.8162$ and 1.8671, and $\upsilon_{\rm r}=11~191$ and $5822$ \kms\ from the quasar system. The \CIV\ NALs with $\upsilon_{\rm r}=5822$ \kms\ did not show continuous variation. The \CIV\ NALs with $\upsilon_{\rm r}=11~191$ \kms\ was continuously weakened from $\rm MJD=53472$ to 57426. A pair of Gaussian functions with $\rm FWHM=566$ \kms\ and 729 \kms\ can well model the absorption features imprinted in the spectra obtained on $\rm MJD=53472$ and 55591, respectively. Although we do not detect absorption features of other transitions at the same redshift, the variable absorption and good fits suggest that the continuous variation absorption feature could be possibly produced by the associated \CIV\ absorber with high velocity.

  \item Quasar SDSS J141334.38+421201.7 with $z_{\rm em}=2.8130$. It was spectroscopically observed by the SDSS on $\rm MJD~=~52823$, 56093 and 57519. In the quasar spectra, we detect 2 \CIV\ NALs with $z_{\rm abs}=2.3968$ and 2.8070, and $\upsilon_{\rm r}=34~521$ and $472$ \kms\ from the quasar system. The system with $\upsilon_{\rm r}=34~521$ was significantly enhanced by $\Delta W_r^{\lambda1548}=3.2\sigma_{\rm \Delta W_r^{\lambda1548}}$ during $\rm MJD=56093$ and 57519.

  \item Quasar SDSS J142500.24+494729.2 with $z_{\rm em}=2.2600$. It was spectroscopically observed by the SDSS on $\rm MJD~=~52460$, 56416 and 57513. In the quasar spectra, we detect 4 \CIV\ NALs with $z_{\rm abs}=1.8425$, 1.8725, 1.9312 and 2.2665, and $\upsilon_{\rm r}=40~857$, $37~751$, $31~774$ and $-597$ \kms\ from the quasar system. The systems with $\upsilon_{\rm r}=40~857$ was weakened first by $\Delta W_r^{\lambda1548}=-2.2\sigma_{\rm \Delta W_r^{\lambda1548}}$ during $\rm MJD=52460$ and 56416, and then enhanced by $\Delta W_r^{\lambda1548}=5.5\sigma_{\rm \Delta W_r^{\lambda1548}}$ from $\rm MJD=56416$ to 57513. Another two systems with $37~751$ and $31~774$ \kms\ were simultaneously enhanced during $\rm MJD=56416$ and 57513 as well.

  \item Quasar SDSS J154857.86+141440.9 with $z_{\rm em}=2.4753$. It was spectroscopically observed by the SDSS on $\rm MJD~=~54570$, 55333 and 55739. In the quasar spectra, we detect 4 \CIV\ NALs with $z_{\rm abs}=2.3127$, 2.3832, 2.4222 and 2.4683, and $\upsilon_{\rm r}=16~364$, $8055$, $4618$ and $604$ \kms\ from the quasar system. The systems with $\upsilon_{\rm r}=16~364$, $8055$ and $4618$ \kms\ were simultaneously varied during the same observation times.

  \item Quasar SDSS J235454.30-092603.2 with $z_{\rm em}=1.9810$. It was spectroscopically observed by the SDSS on $\rm MJD~=~52201$, 56602 and 56604. In the quasar spectra, we detect 4 \CIV\ NALs with $z_{\rm abs}=1.6278$, 1.8179, 1.8786 and 1.9169, and $\upsilon_{\rm r}=37~634$, $16~862$, $10~482$ and $6520$ \kms\ from the quasar system. The system with $\upsilon_{\rm r}=16~862$ \kms\ was stable between $\rm MJD=52201$ and 56602, while it was significantly enhanced by $\Delta W_r^{\lambda1548}=2.7\sigma_{\rm \Delta W_r^{\lambda1548}}$ during $\rm MJD=56602$ and 56604. The other one variable system with $\upsilon_{\rm r}=10~482$ \kms\ was also slightly enhanced by $\Delta W_r^{\lambda1548}=2.1\sigma_{\rm \Delta W_r^{\lambda1548}}$ from $\rm MJD=52201$ to 56604.
\end{enumerate}

\subsection{Correlated variations of absorption lines and continuum}
\label{sect:Correlated_variation}
Both the fluctuations of incident photons and the absorber motions can give rise to variable NALs. The scheme of absorber motion cannot well explain the coherent variation of multiple absorption line systems, which requires a coordinated motion between multiple absorbers. In addition, the initially enhanced and then weakened systems, or the opposite cases, are very difficultly to explain using the absorber motions. For example, the \CIV\ system with $z_{\rm abs}=2.3134$, which was imprinted in the spectra of quasar SDSS J084525.84+072222.3, was enhanced first by $\Delta W_r^{\lambda1548}=2.9\sigma_{\rm \Delta W_r^{\lambda1548}}$ from $\rm MJD=52964$ to 55893, and then was weakened by $\Delta W_r^{\lambda1548}=-3.1\sigma_{\rm \Delta W_r^{\lambda1548}}$ during $\rm MJD=55893$ and 55964. These complex variations require an absorber moving first into and then out of the quasar sightline, in the scheme of the absorber motion driving variable NALs. Nevertheless, fluctuations of incident photons can reasonably interpret the coherence variations of multiple absorption line systems and the complex variations of a single system.

\begin{figure}
\centering
\includegraphics[width=0.45\textwidth]{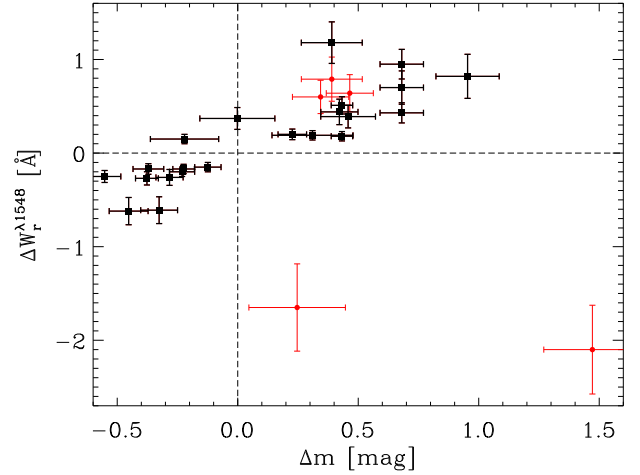}
\caption{Variations of the \CIV\ NALs with $\mid\Delta W_r^{\lambda1548}\mid\ge3\sigma_{\rm \Delta W_r^{\lambda1548}}$ against those of the continuum fluxes at 1350 \AA. The red symbols represent the disappeared or emergent \CIV\ NALs. $\Delta W_{\rm r}^{\lambda1548} >0$ represents enhanced \CIV\ NALs, and $\Delta W_{\rm r}^{\lambda1548} <0$ is the opposite cases. $\Delta m<0$ represents brightened objects, and $\Delta m>0$ represents the opposite cases. The vertical and horizontal dashed lines indicate the 0 positions.}
\label{fig:delta_w_m}
\end{figure}

Among the 19 obviously variable \CIV\ NAL systems described in Section \ref{sect:variable_NALs}, 13 systems are the cases in which multiple systems change together. In terms of the above discussions, the variations of these \CIV\ NALs were likely caused by the fluctuations of the incident continuum. Changes in the incident photons into the absorbing clouds can be caused by the intrinsic variation of the background ionizing continuum, and by the variation of the column density or ionization of the shielding gas located between the absorbing clouds and the ionizing continuum source \cite[e.g.,][]{1995ApJ...451..498M,2007ApJ...660..152M}. To further investigate the origins of the variations of \CIV\ NALs, we explore the relationship between the variations of the quasar continuum and the \CIV\ NALs. Out results are shown in Figure \ref{fig:delta_w_m}, which clearly exhibits a tight correlation between the variations of the 1350 \AA\ continuum and the \CIV\ NALs. For all the 27 \CIV\ NAL pairs with $\mid\Delta W_r^{\lambda1548}\mid\ge3\sigma_{\rm \Delta W_r^{\lambda1548}}$ (the black and red symbols in Figure \ref{fig:delta_w_m}), the Spearman's correlation test yields a correlation coefficient $\rho=0.63$ and a probability $P < 10^{-3}$ of no correlation. When excluding the disappeared and emergent \CIV\ systems (the red symbols in Figure \ref{fig:delta_w_m}), the Spearman's correlation test suggests a stronger correlation ($\rho=0.89$ and $P<10^{-8}$). This tight correlation is consistent with the results reported by previous works \cite[e.g.,][]{2015MNRAS.454.3962H,2017ApJS..229...22H,2018MNRAS.473L.106L,2018MNRAS.474.3397L}. The increase or decrease of the column density of the shielding gas would not result in an obviously correlated variation between the ionizing continuum and the absorption lines. In addition, the variations of the ionization of the shield gas, which are introduced by the alterations in the ionization state, indeed reflect the changes in the ionizing continuum. Therefore, combining with the coordinated variations of multiple \CIV\ systems, the significant correlation between the variations of the continuum fluxes and the absorption strengths of \CIV\ NALs suggests that the variations of \CIV\ NALs would be mainly driven by the changes of the ionizing continuum.

Except for one enhanced \CIV\ NAL shown in the top left corner and two disappeared \CIV\ NALs shown in the bottom right corner, Figure \ref{fig:delta_w_m} indicates that the variations of absorption lines are inversely related to the changes of the ionizing continuum, namely, the brighter objects the weaker absorption lines. Figure 11 of \cite{2015ApJ...814..150W} demonstrated that the responses of the ionic column density to the changes in the ionizing continuum are connected to the ionization state of the absorbing gas. Below the critical value of the ionization parameter, the fraction of the $\rm C^{3+}$ increases as the ionization parameter increases, and the variations of the \CIV\ absorption lines positively respond to the changes in the ionizing continuum. Nevertheless, they are oppositive cases when the ionization parameter is larger than the critical value. The variations of the \CIV\ NALs shown in \ref{fig:delta_w_m} are inversely correlated with the continuum fluxes, which implies that our 19 obviously variable \CIV\ systems may be dominated by the gas with high ionization phase.

\subsection{Relative velocity and timescales of variable absorption line systems}
\label{sect:timescale_variation}
\begin{figure}
\centering
\includegraphics[width=0.45\textwidth]{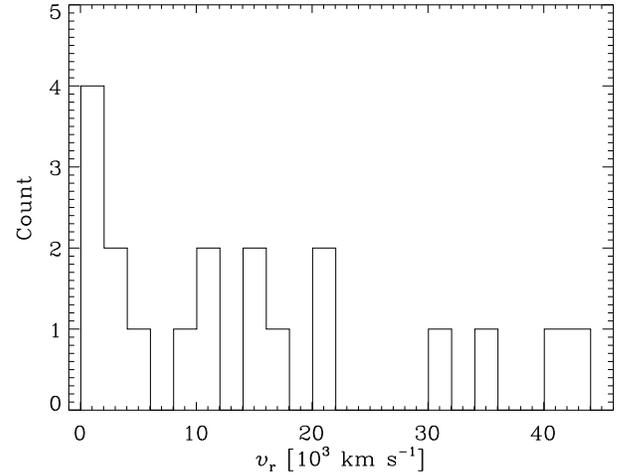}
\caption{Relative velocity distribution of the 19 obviously variable \CIV\ narrow absorption line systems.}
\label{fig:vr}
\end{figure}

Figure \ref{fig:vr} shows the relative velocities of the 19 obviously variable \CIV\ NALs. In general, associated NALs are limited to $\upsilon_{\rm}\le5000$ \kms\ \cite[e.g.,][]{2004ApJ...613..129W}, while here we find that there are 14 systems with $\upsilon_{\rm}>0.01c$, and 4 systems with $\upsilon_{\rm}>0.1c$, where $c$ is the speed of light. This suggests that some of our \CIV\ NALs might be related to quasar outflows with high velocities.

\begin{figure*}
\centering
\includegraphics[width=0.31\textwidth]{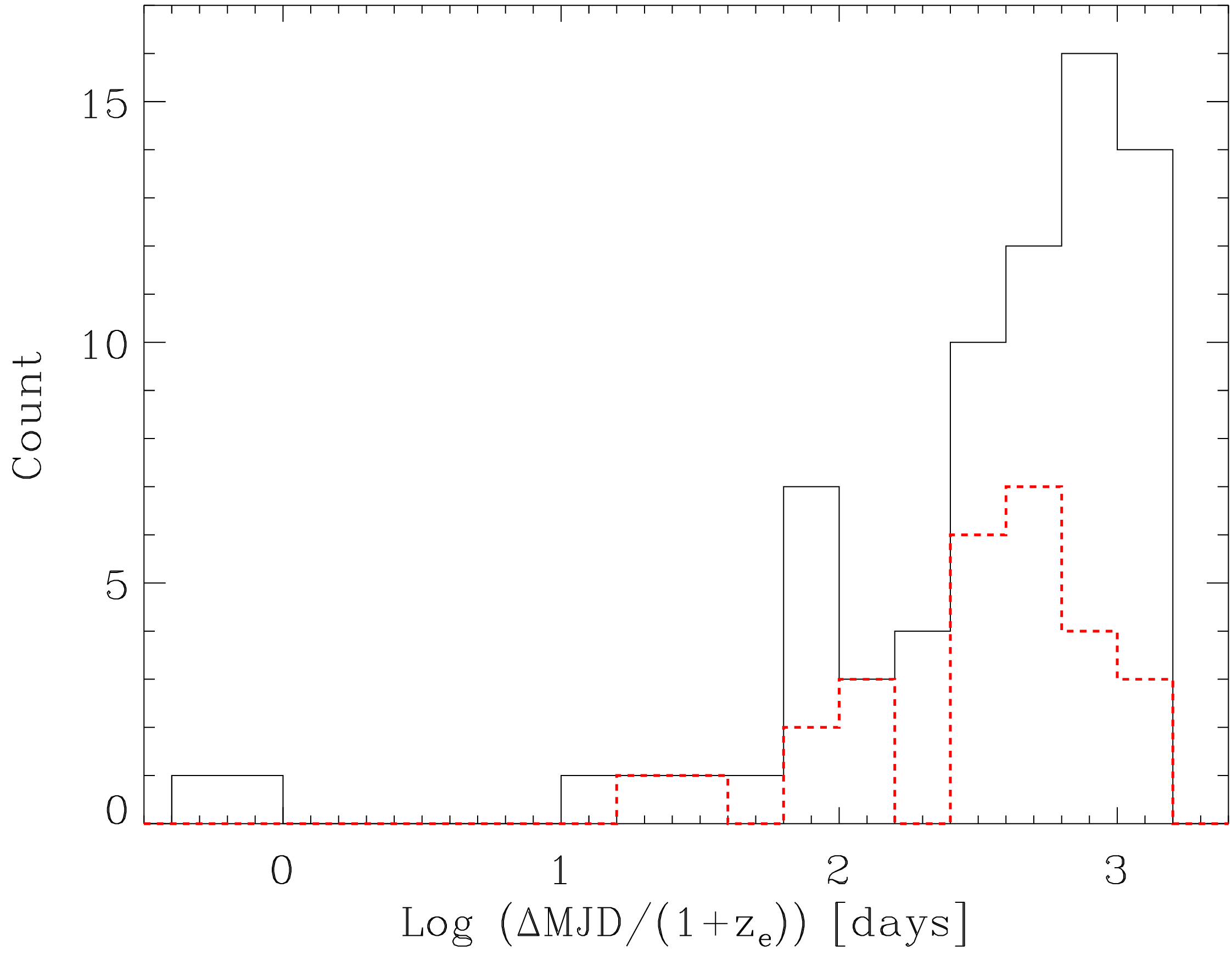}
\hspace{2ex}
\includegraphics[width=0.31\textwidth]{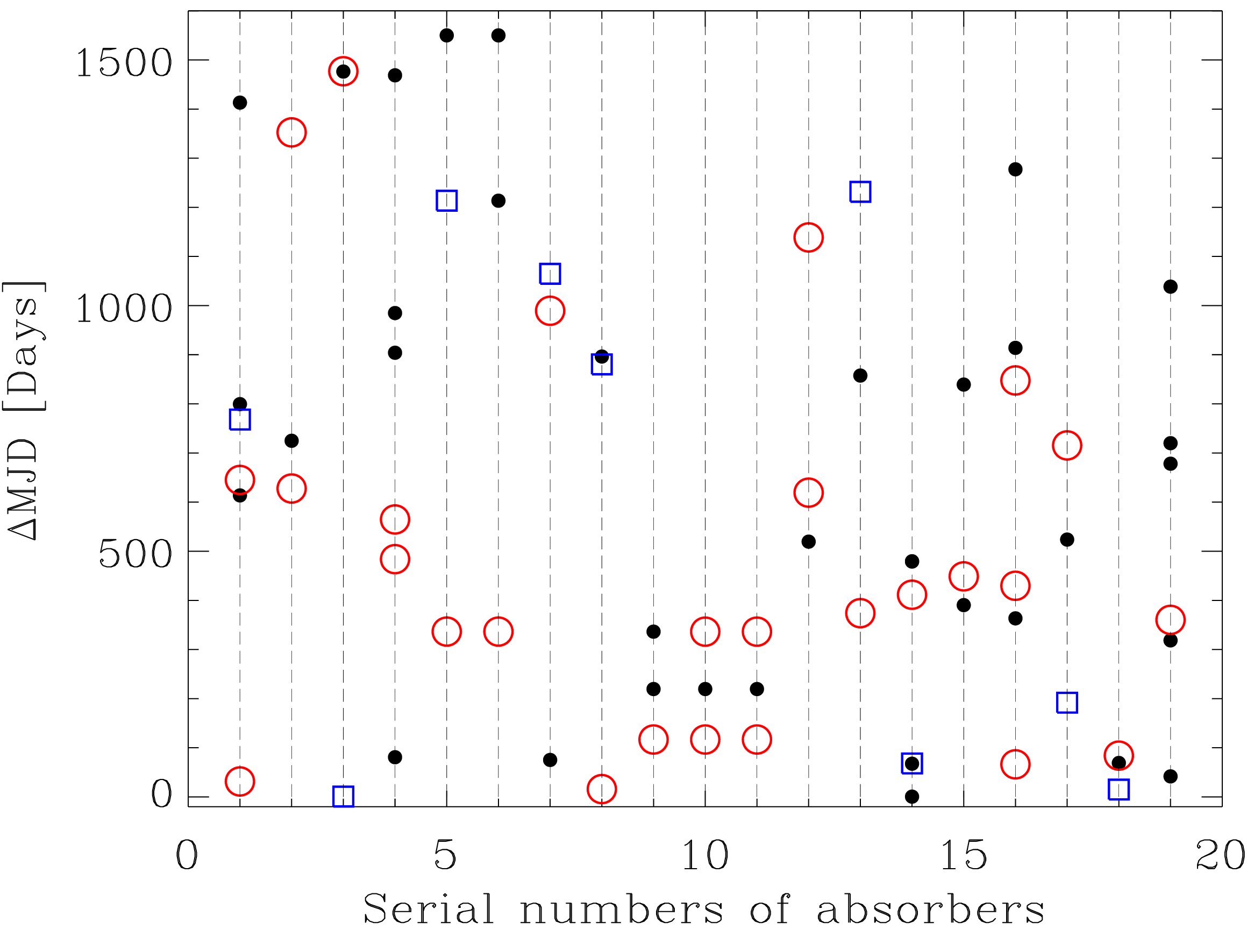}
\hspace{2ex}
\includegraphics[width=0.31\textwidth]{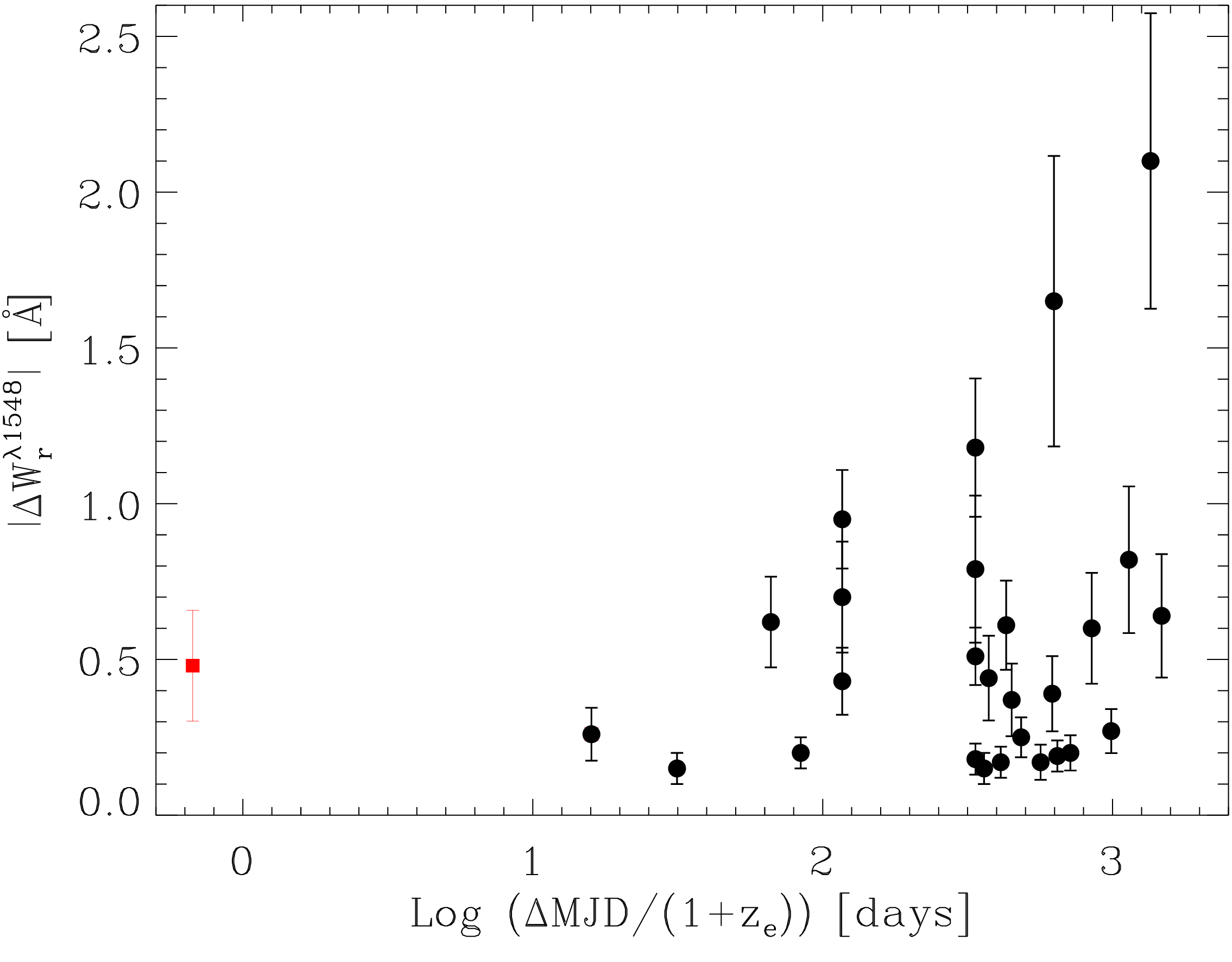}
\caption{Distributions of time intervals and $\mid\Delta W_r^{\lambda1548}\mid$ of the 19 obviously variable \CIV\ NALs. The left panel: the black solid line represents the time intervals of the all 72 \CIV\ NAL-pairs, and the red dash line only represents the time intervals of the 27 \CIV\ NAL-pairs with $\mid\Delta W_r^{\lambda1548}\mid\ge3\sigma_{\rm \Delta W_r^{\lambda1548}}$. The middle panel: the x-axis indicates the serial numbers of the 19 obviously variable \CIV\ NALs, the red open circles represent the NAL-pairs with $\mid\Delta W_r^{\lambda1548}\mid\ge3\sigma_{\rm \Delta W_r^{\lambda1548}}$, the black filled circles represent the NAL-pairs with $\mid\Delta W_r^{\lambda1548}\mid<2\sigma_{\rm \Delta W_r^{\lambda1548}}$, and the blue unfilled squares represent the NAL-pairs with $2\sigma_{\rm \Delta W_r^{\lambda1548}}\le\mid\Delta W_r^{\lambda1548}\mid<3\sigma_{\rm \Delta W_r^{\lambda1548}}$. The right panel: changes in $W_{\rm r}^{\lambda1548}$ against time intervals. The black circles represent the \CIV\ NALs with $\mid\Delta W_r^{\lambda1548}\mid\ge3\sigma_{\rm \Delta W_r^{\lambda1548}}$, and the red square is the \CIV\ NALs that was enhanced by $\Delta W_r^{\lambda1548}=2.7\sigma_{\rm \Delta W_r^{\lambda1548}}$ in a very short time of 0.67 day at rest frame.}
\label{fig:delta_mjd_variableabs}
\end{figure*}

Coordinated variations of the ionizing continuum and absorption lines require a recombination time of recombing gas that is shorter than the timescale of continuum fluctuations and the time interval between two observations \cite[e.g.,][]{1992ApJ...397...81B}; otherwise, one cannot observe obvious variations in the absorption lines. The time intervals of the 72 \CIV\ NAL pairs of the 19 obviously variable \CIV\ NALs are from $\rm \Delta MJD=0.52$ to 1550 days at rest-frame, these intervals are shown in the left and middle panels of Figure \ref{fig:delta_mjd_variableabs}. We note that a \CIV\ system with $\upsilon_{\rm r}=16~862$ \kms, which was imprinted in the spectra of the quasar SDSS J235454.30-092603.2, was enhanced by $\Delta W_r^{\lambda1548}=2.7\sigma_{\rm \Delta W_r^{\lambda1548}}$ on a very short time interval of 0.67 day at rest-frame.

Although there is not a tightly linear correlation between changes in absorption lines and time intervals, it has been discovered for both the NALs \cite[e.g.,][]{2013MNRAS.434..163H,2015MNRAS.450.3904C} and BALs \cite[e.g.,][]{2013ApJ...777..168F,2014ApJ...792...77M} that the variations of absorption lines depend on the time intervals. We show the changes in $W_{\rm r}^{\lambda1548}$ of \CIV\ NALs in the right panel of Figure \ref{fig:delta_mjd_variableabs}, which clearly indicates that a large variation of absorption lines is very difficultly to form on a short time interval though there is a \CIV\ NALs enhanced by $\Delta W_r^{\lambda1548}=2.7\sigma_{\rm \Delta W_r^{\lambda1548}}$ in 0.67 day at rest-frame. In addition, the middle panel of Figure \ref{fig:delta_mjd_variableabs} shows that in the same systems, the obvious variations are not positively related to the time intervals.

\section{Summary}
\label{sect:summary}
Using the SDSS spectra of 207 quasars with at least three observations and the 1350 \AA\ continuum variations $\Delta f_{\rm 1350}\ge4\sigma_{\rm f_{1350}}$, we analyze the \CIV\ NALs and investigate their variations. Our results and conclusions are as follows.
\begin{itemize}
  \item There are 166 quasars for which at least 1 \CIV\ NAL system with $W_{\rm r}^{\lambda1548}\ge3\sigma_{\rm W_{\rm r}^{\lambda1548}}$ and $W_{\rm r}^{\lambda1551}\ge2\sigma_{\rm W_{\rm r}^{\lambda1551}}$ was detected in one of the multi-epoch spectra. We obtain 328 \CIV\ NAL systems, comprising 1210 \CIV\ NAL pairs. We find that 19 out of 328 \CIV\ NAL systems were changed by $\mid\Delta W_r^{\lambda1548}\mid\ge3\sigma_{\rm \Delta W_r^{\lambda1548}}$. Among these 19 obviously variable \CIV\ systems, 13 systems are accompanied by other variable \CIV\ systems, and 9 systems were changed continuously during multiple observations. In addition, 4 out of 19 systems have $\upsilon_{\rm r}>0.1c$, which are significantly larger than the boundary of 5000 \kms\ that is generally used to define associated \CIV\ NALs. This suggests that some of our \CIV\ NALs might be related to quasar outflows with high velocities.
  \item The variations of absorption lines are tightly and inversely correlated with the changes in the ionizing continuum. Therefore, our 19 obviously variable \CIV\ systems negatively respond to the fluctuations of the ionizing continuum, and might be dominated by gas with high ionization phases.
  \item We find that a \CIV\ system with $\upsilon_{\rm r}=16~862$ \kms, which was imprinted in the spectra of quasar SDSS J235454.30-092603.2, was enhanced by $\Delta W_r^{\lambda1548}=2.7\sigma_{\rm \Delta W_r^{\lambda1548}}$ on a very short time interval of 0.67 day at rest-frame.
  \item Large variations of \CIV\ NALs are difficultly formed in a short timescale.
\end{itemize}

\acknowledgements We much thank the anonymous referee for very helpful comments. This work was supported by the National Natural Science Foundation of China (No. 11763001; No. 11363001; No. 11661012), and the Guangxi Natural Science Foundation (2015GXNSFBA139004).

\appendix
\section{The spectra of the variable \CIV\ absorption systems}

\begin{figure*}
\centering
\includegraphics[width=5.8cm,height=3.8cm]{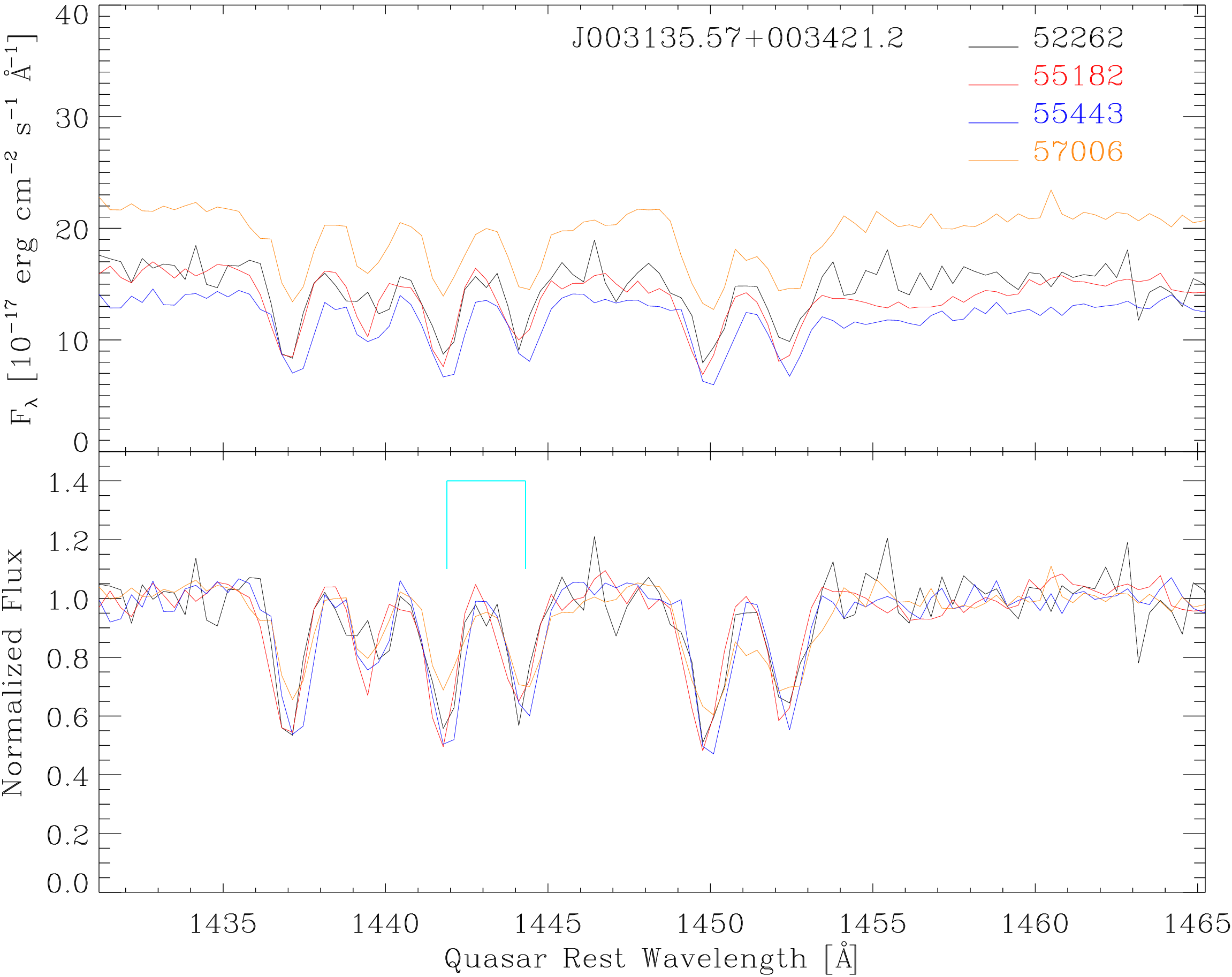}
\hspace{1ex}
\includegraphics[width=5.8cm,height=3.8cm]{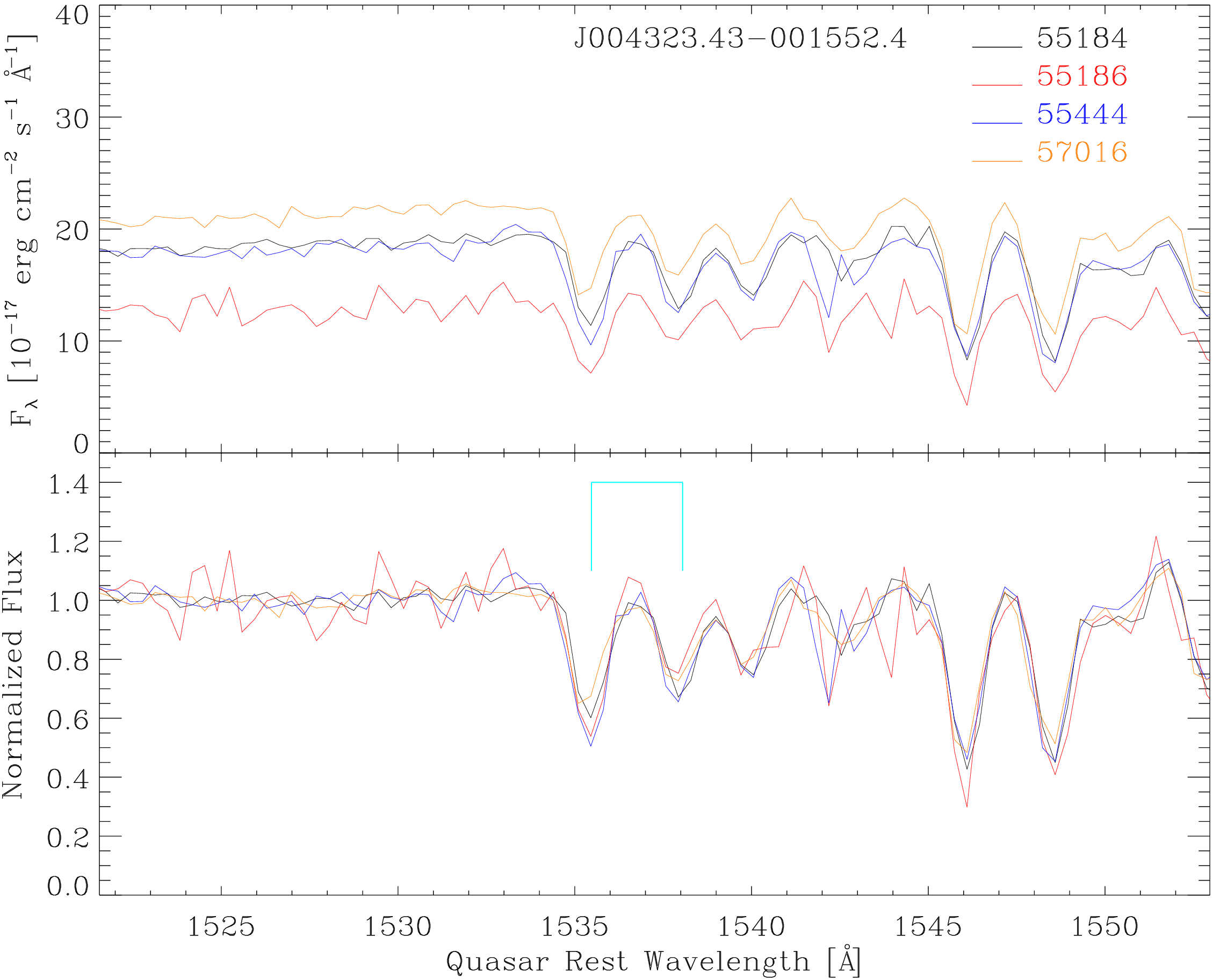}
\hspace{1ex}\vspace{1ex}
\includegraphics[width=5.8cm,height=3.8cm]{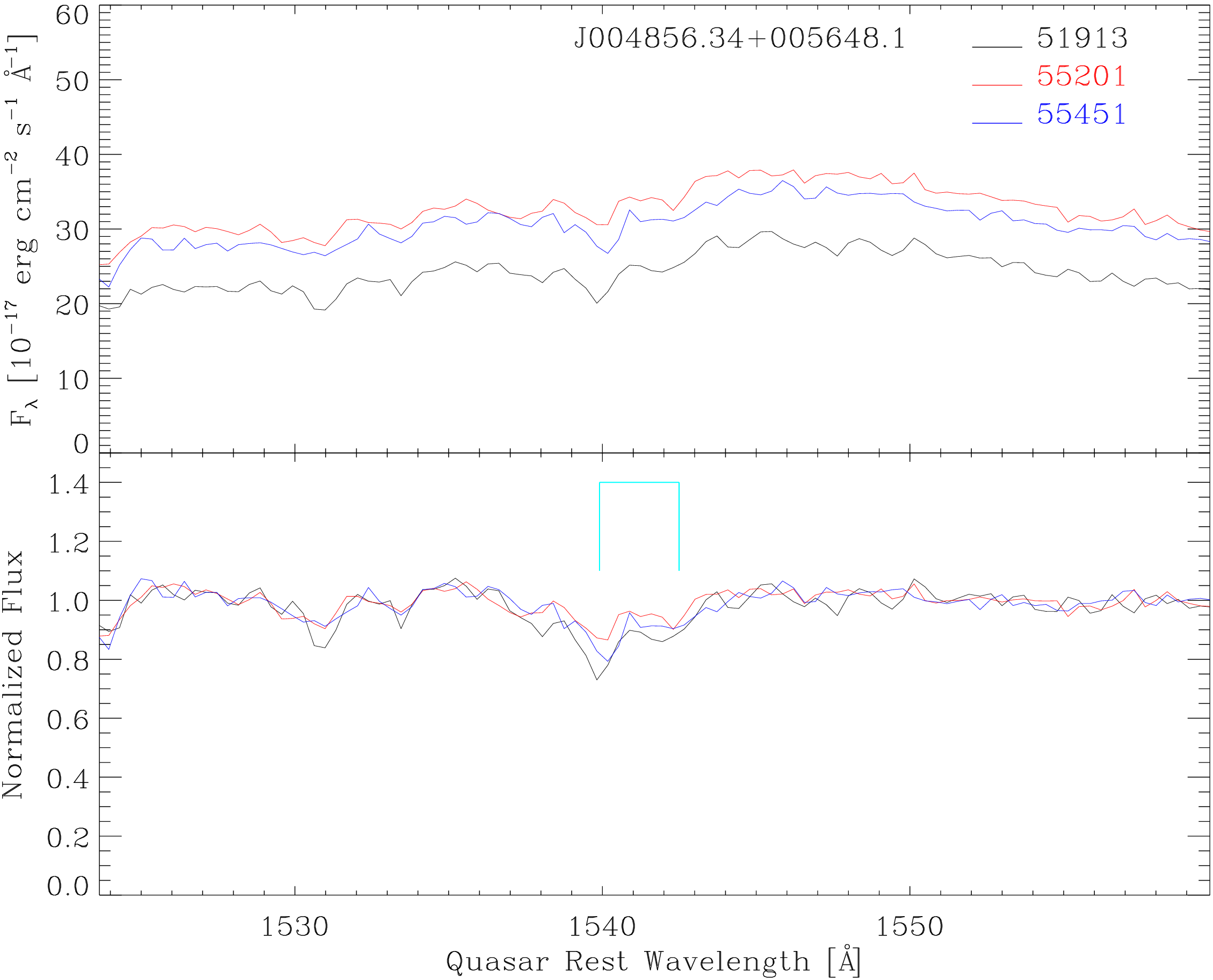}
\includegraphics[width=5.8cm,height=3.8cm]{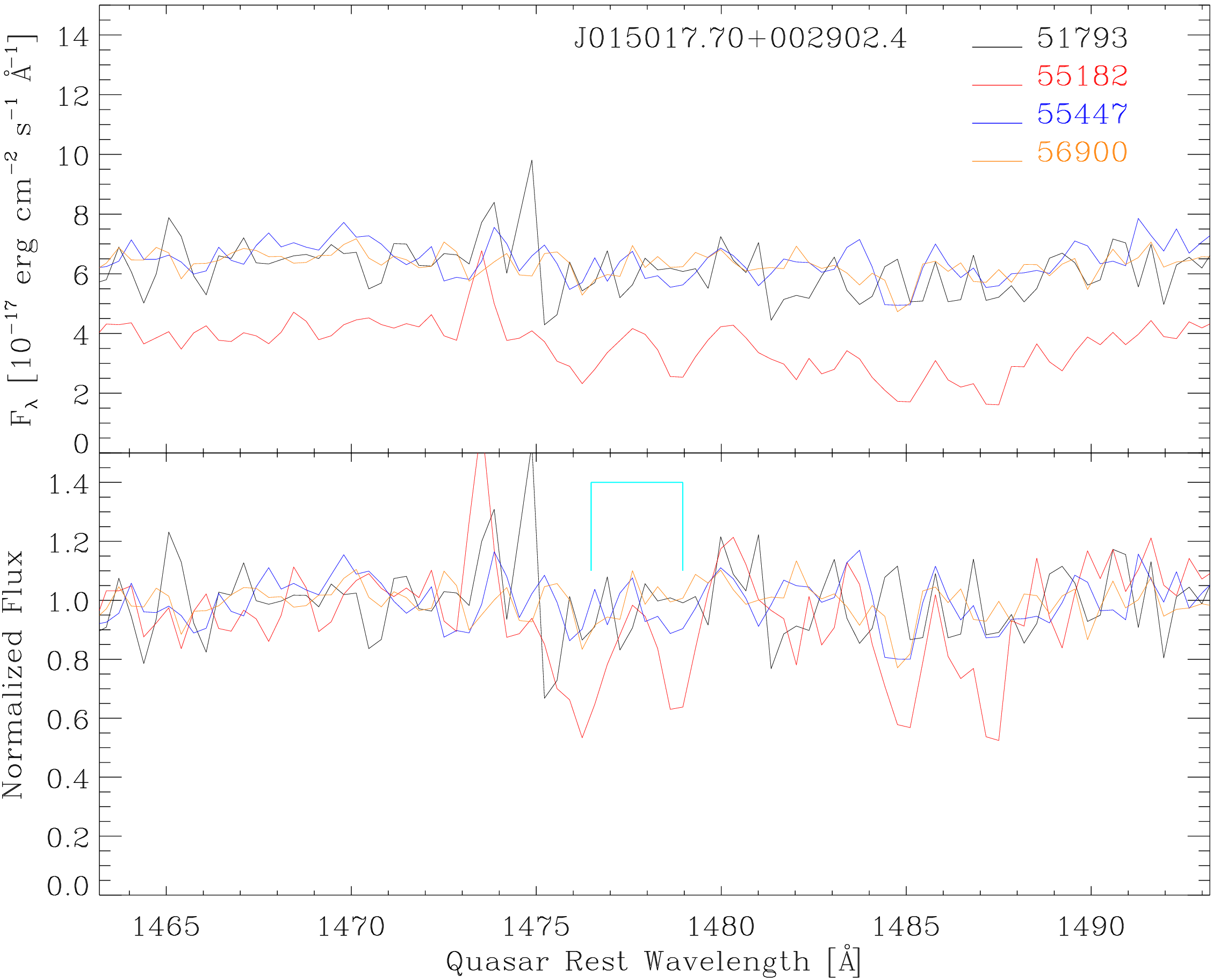}
\hspace{1ex}
\includegraphics[width=5.8cm,height=3.8cm]{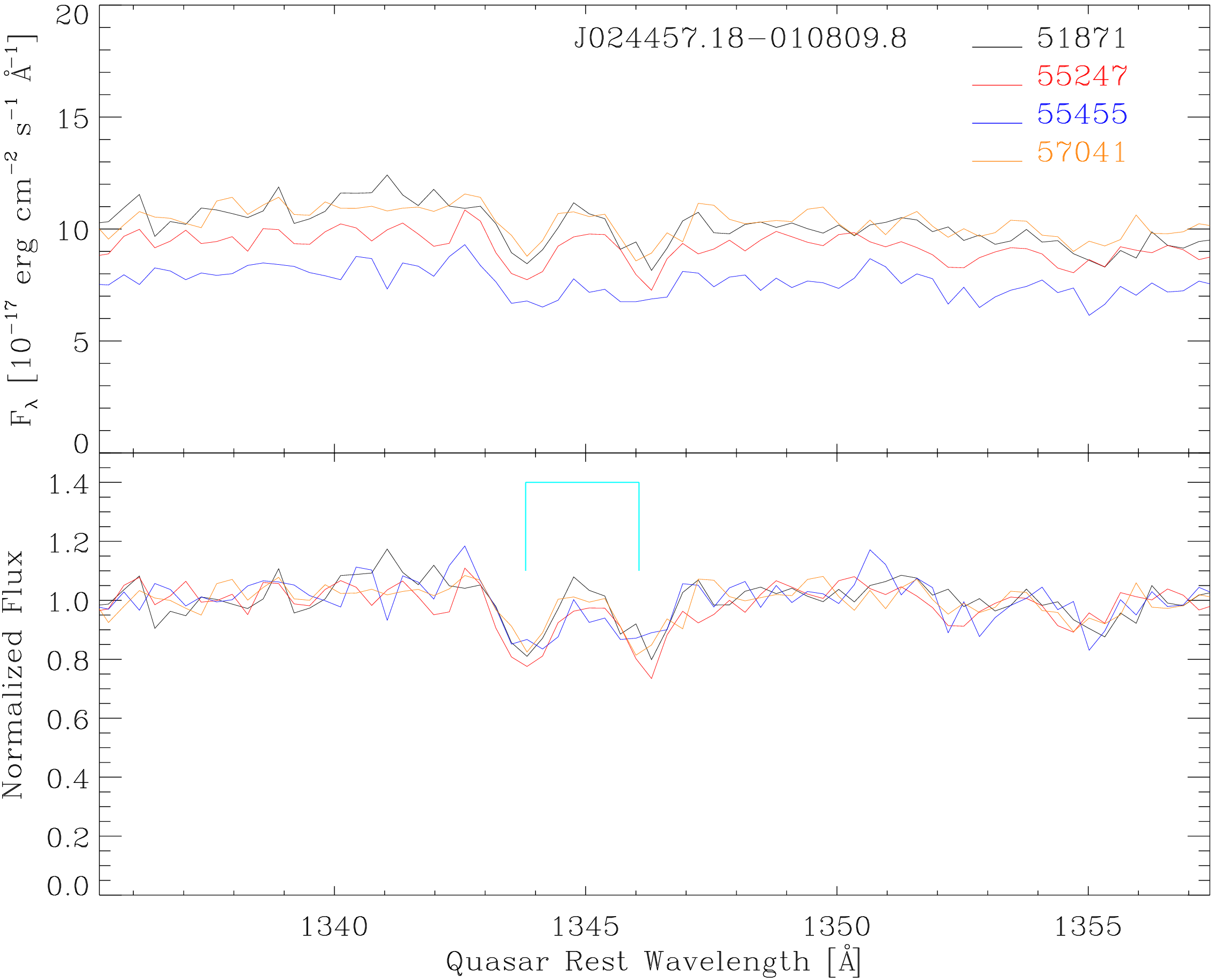}
\hspace{1ex}\vspace{1ex}
\includegraphics[width=5.8cm,height=3.8cm]{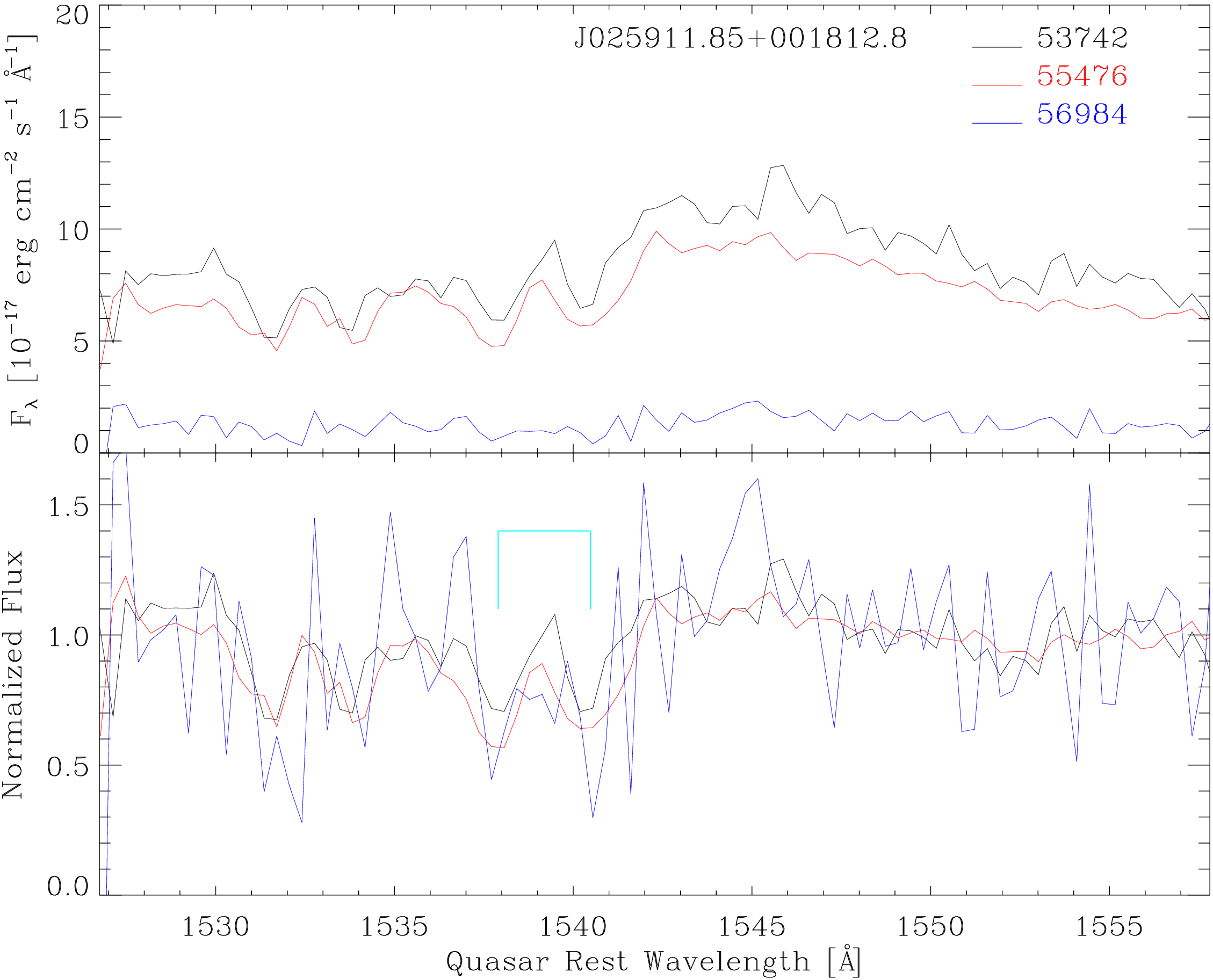}
\hspace{1ex}\vspace{1ex}
\includegraphics[width=5.8cm,height=3.8cm]{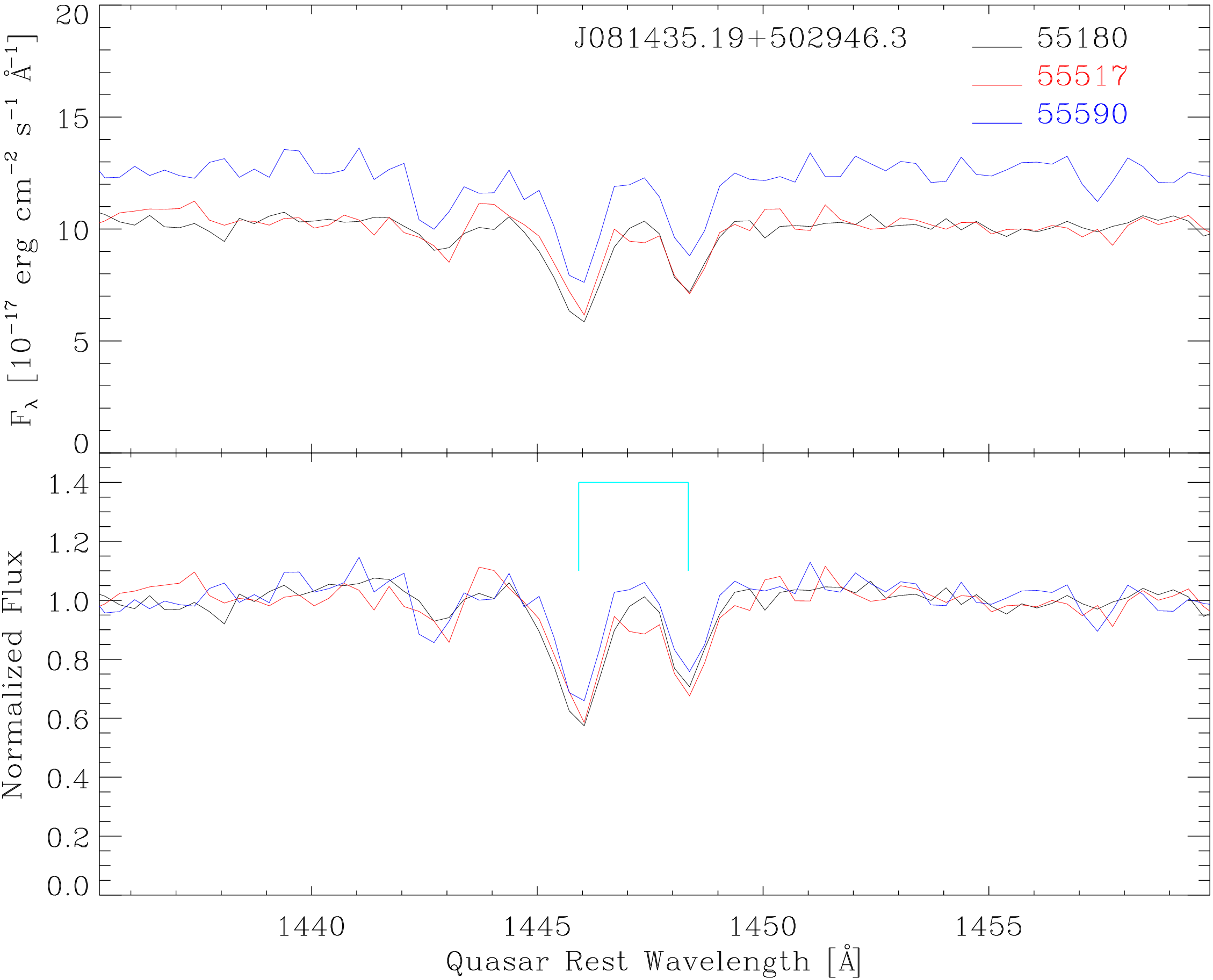}
\includegraphics[width=5.8cm,height=3.8cm]{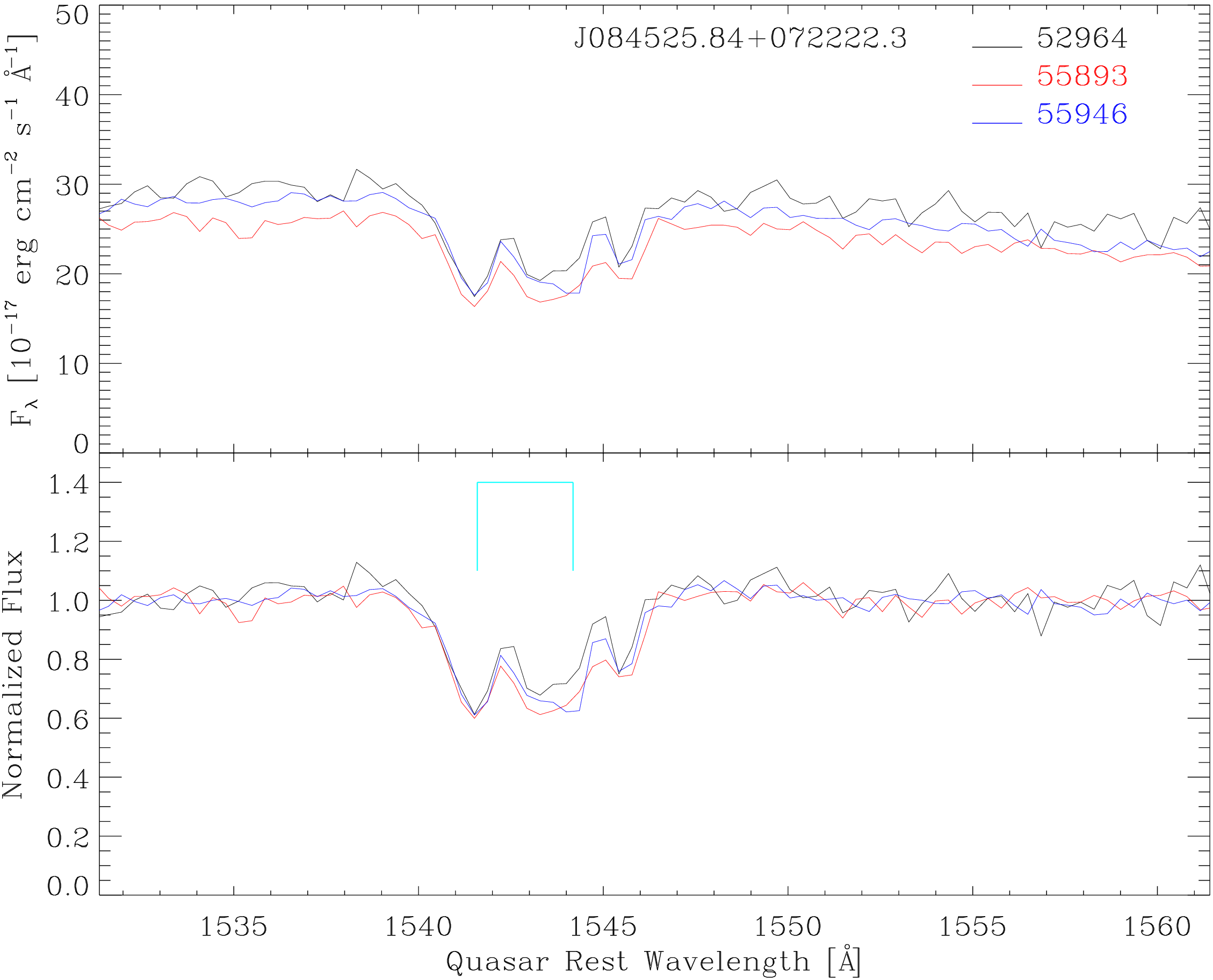}
\hspace{1ex}
\includegraphics[width=5.8cm,height=3.8cm]{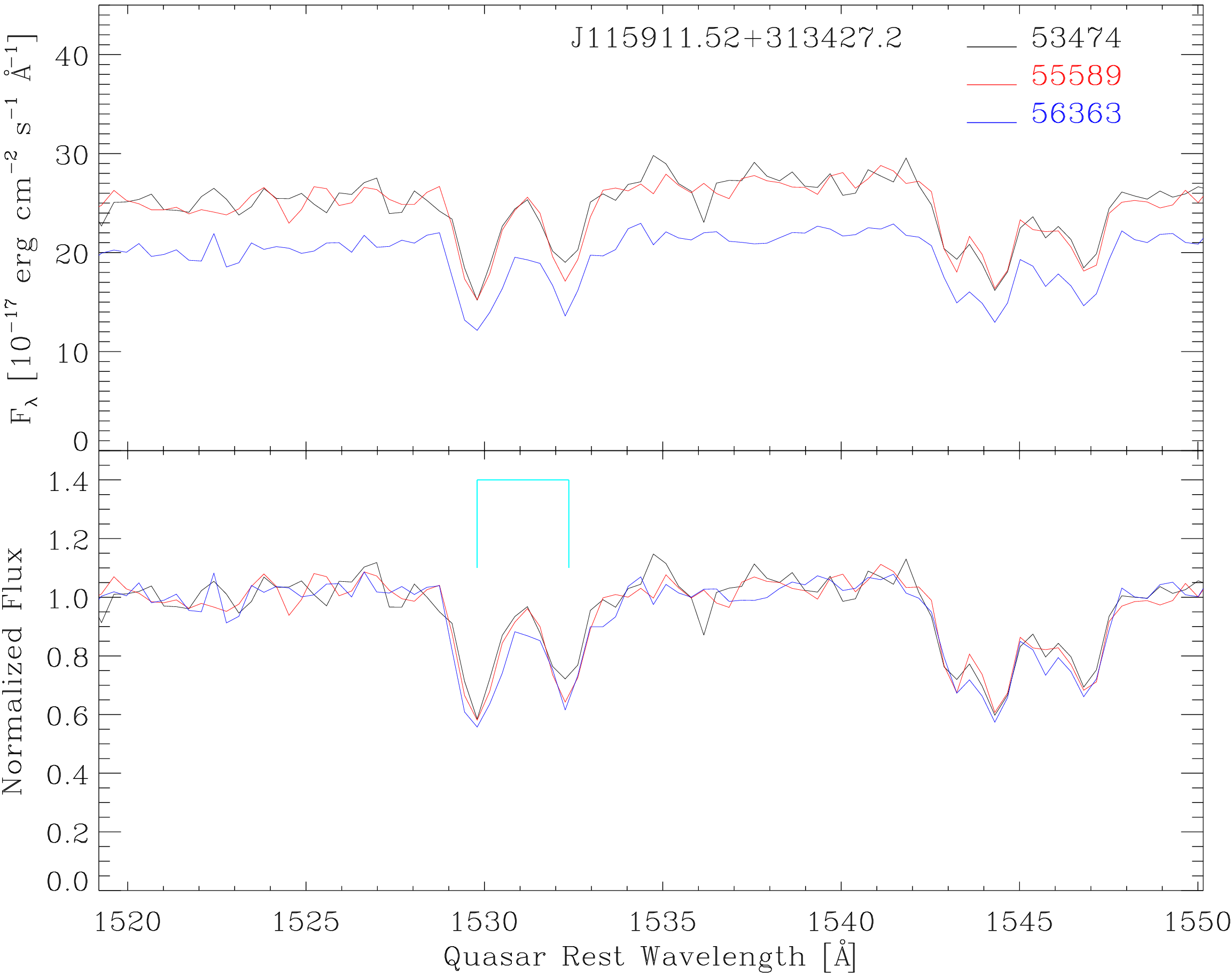}
\hspace{1ex}
\includegraphics[width=5.8cm,height=3.8cm]{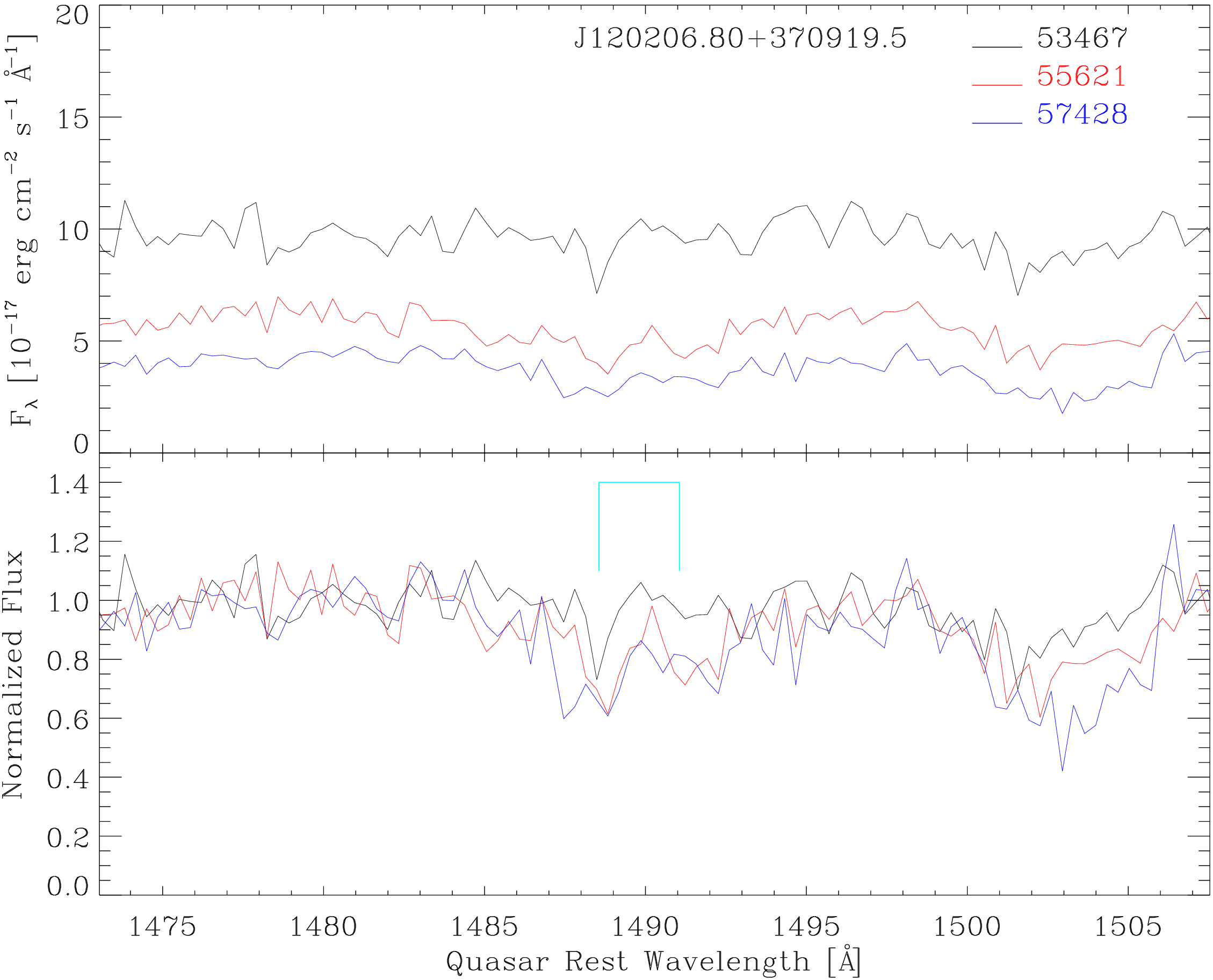}
\hspace{1ex}\vspace{1ex}
\includegraphics[width=5.8cm,height=3.8cm]{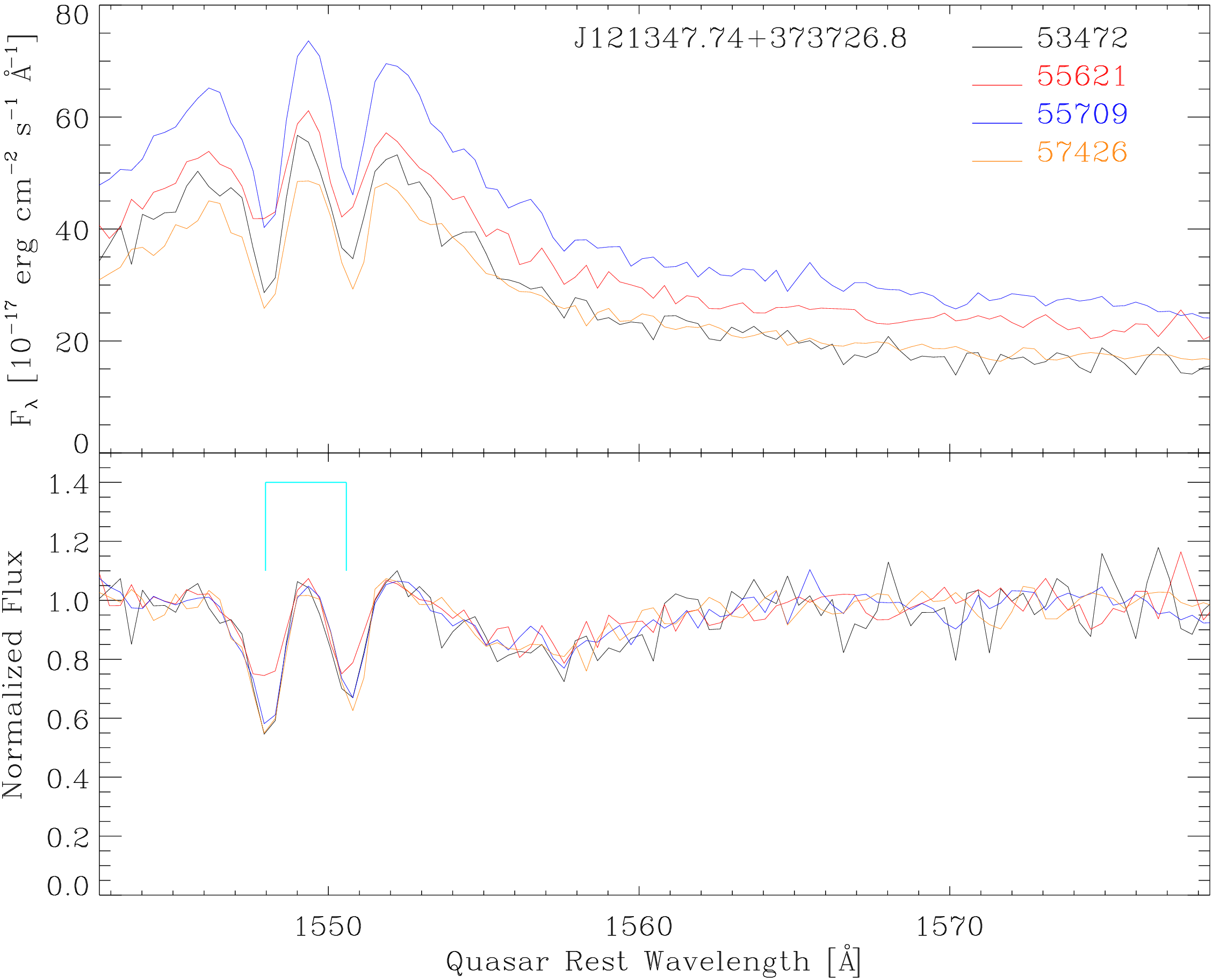}
\includegraphics[width=5.8cm,height=3.8cm]{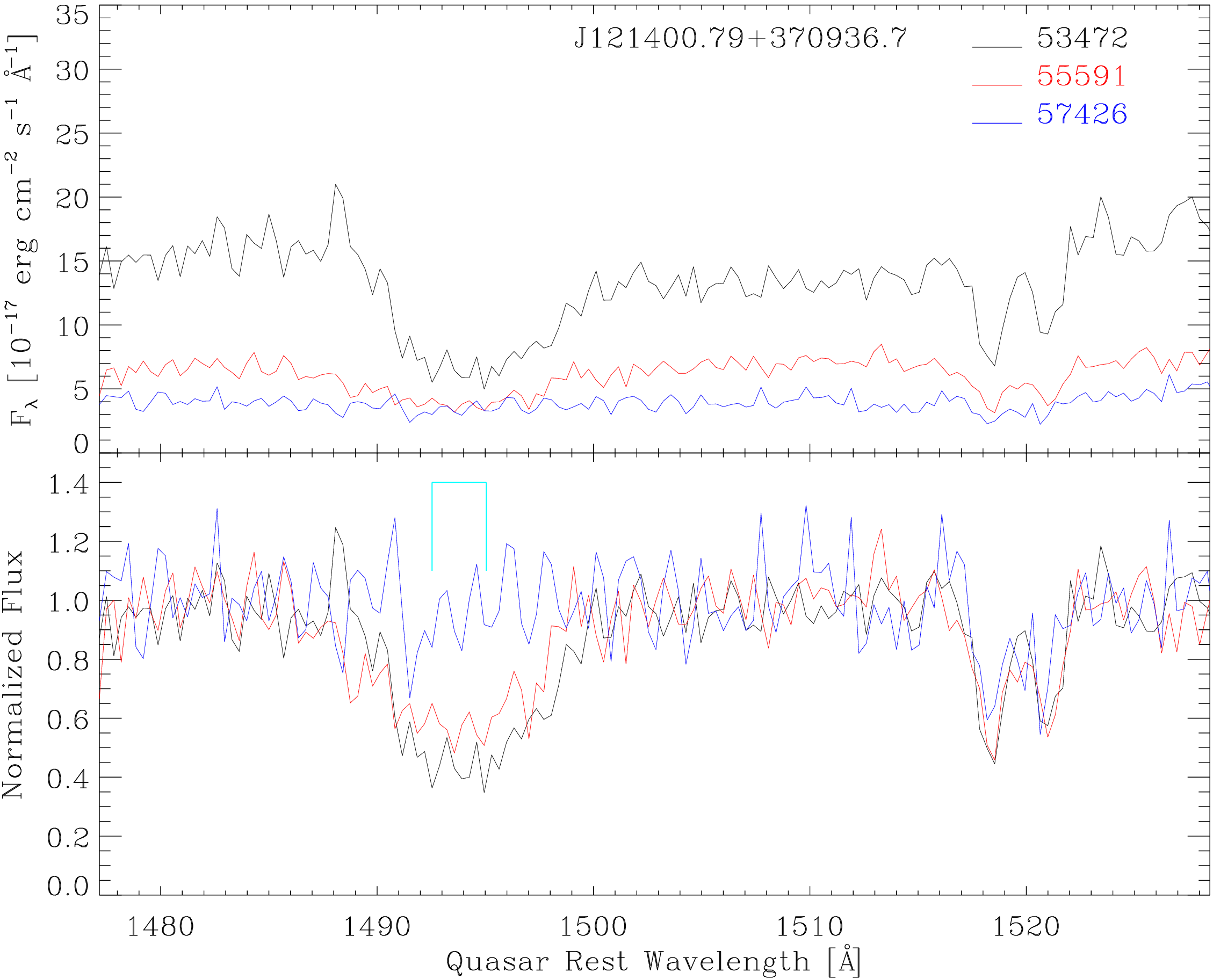}
\hspace{1ex}
\includegraphics[width=5.8cm,height=3.8cm]{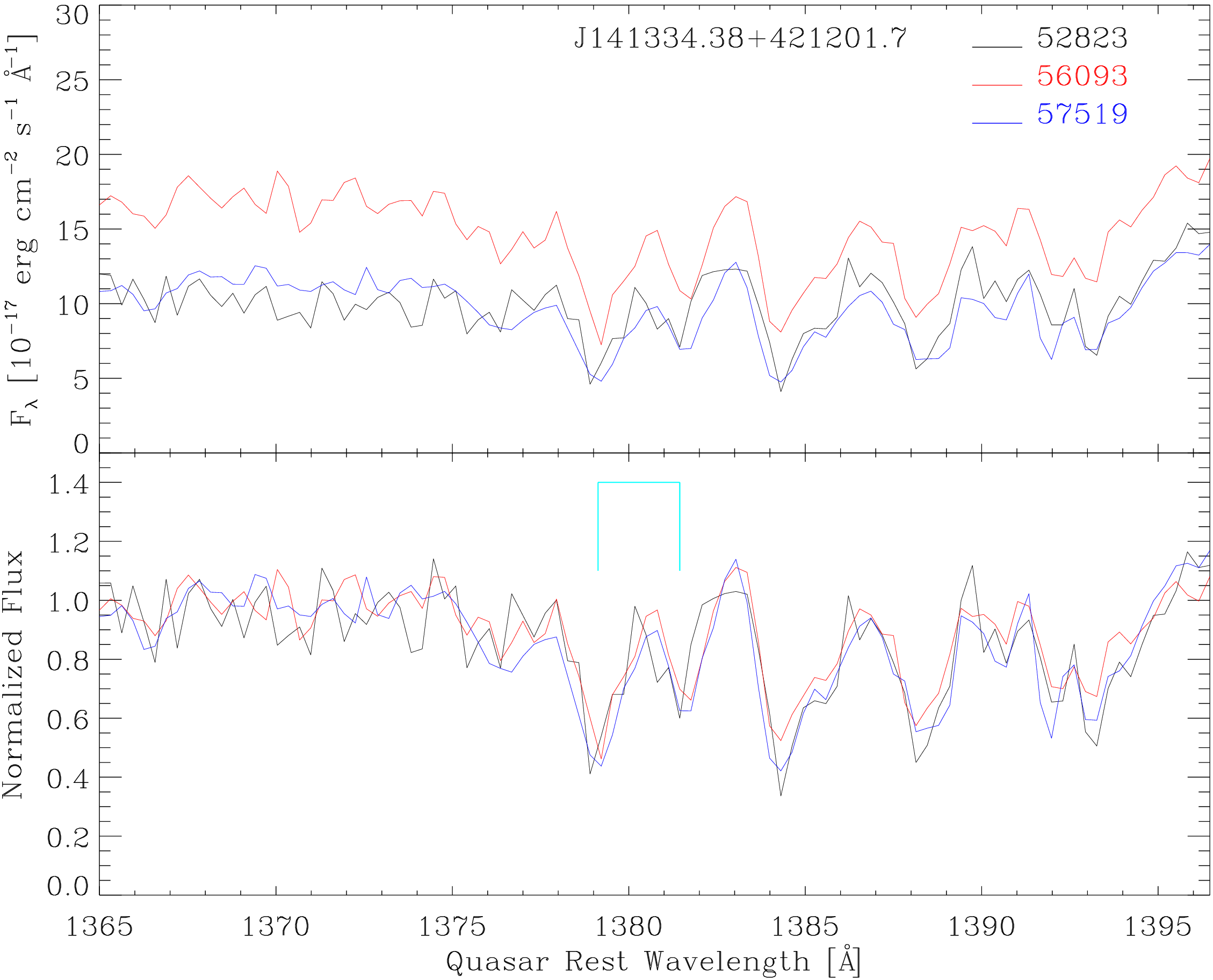}
\hspace{1ex}
\includegraphics[width=5.8cm,height=3.8cm]{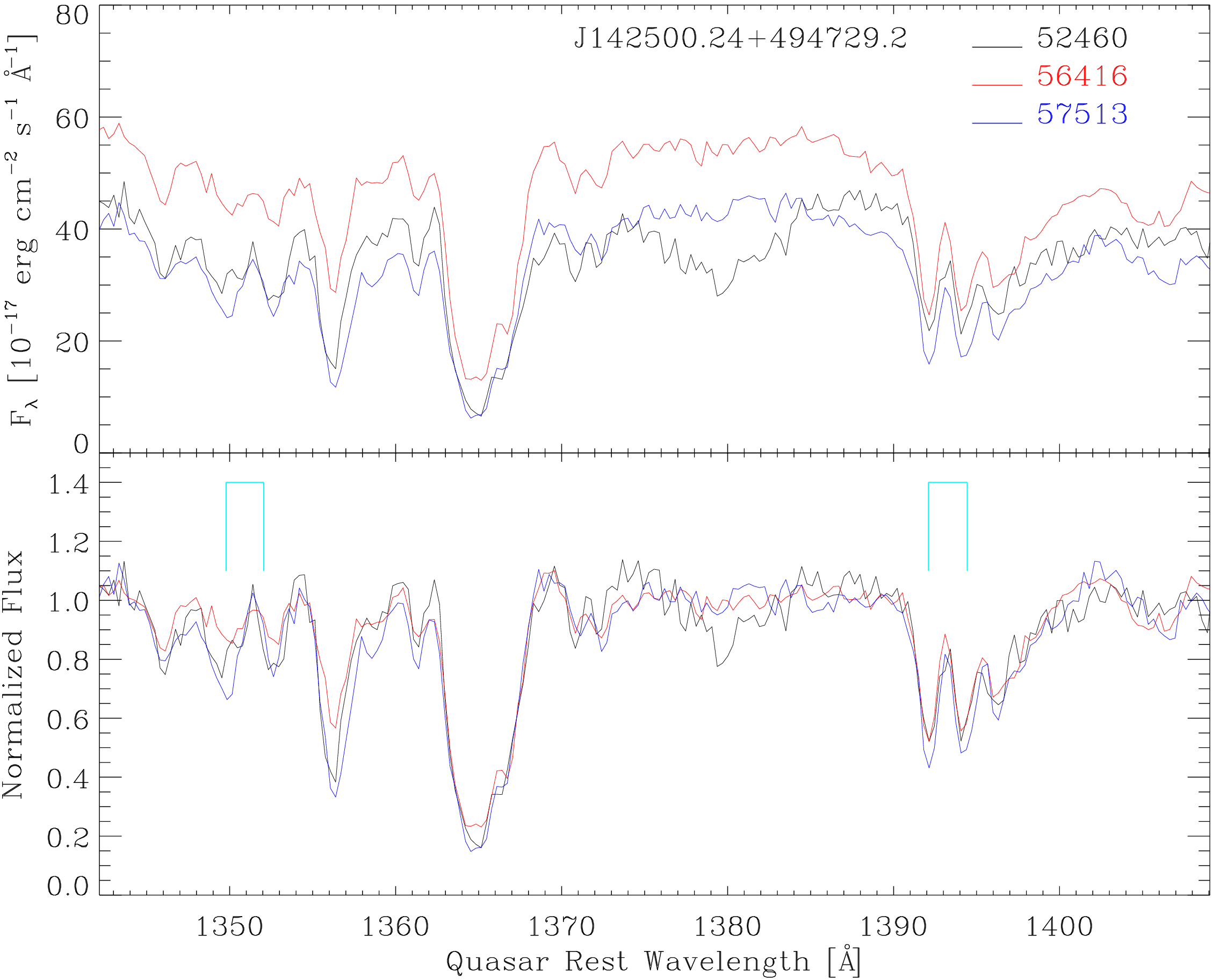}
\hspace{1ex}\vspace{1ex}
\includegraphics[width=5.8cm,height=3.8cm]{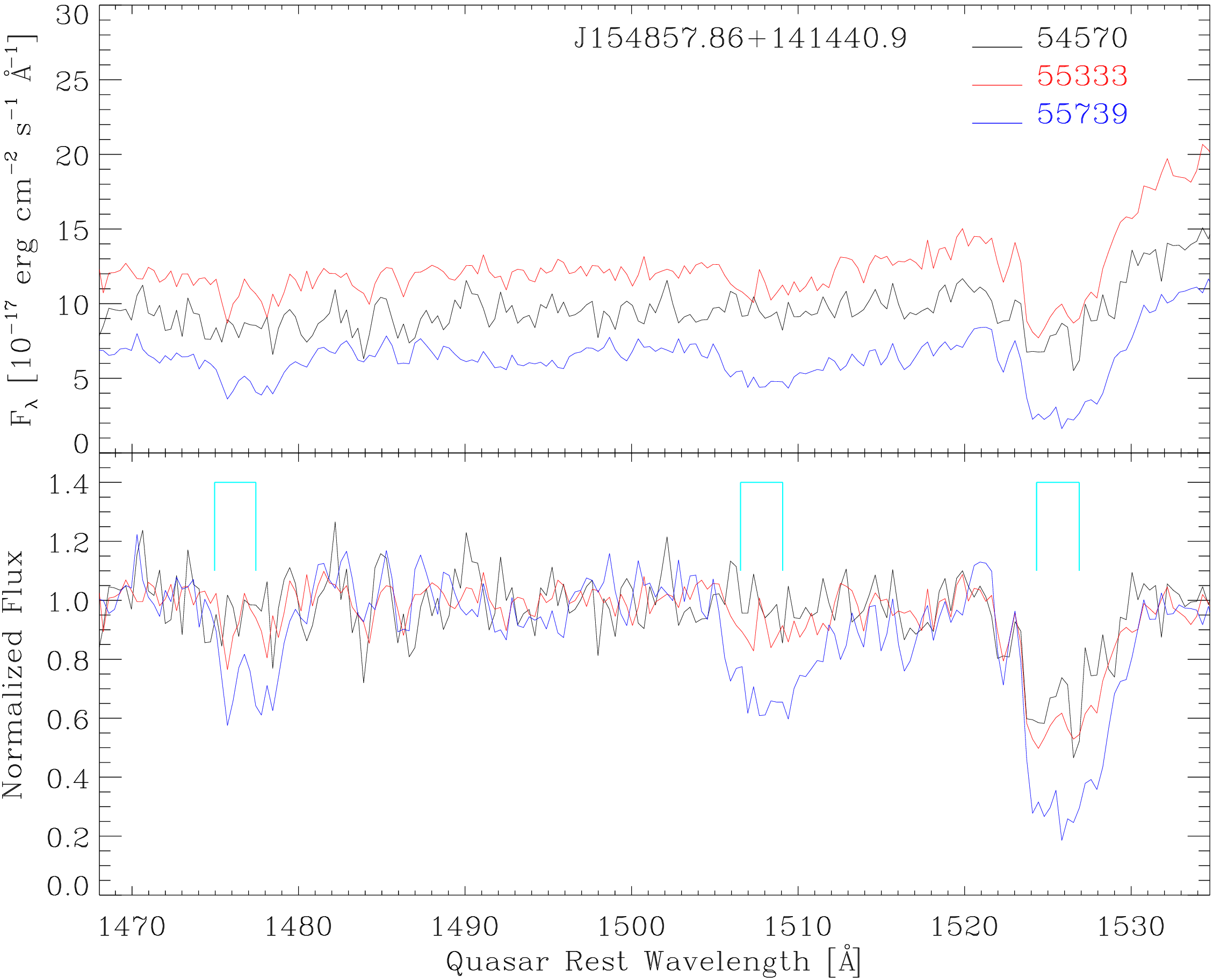}
\includegraphics[width=5.8cm,height=3.8cm]{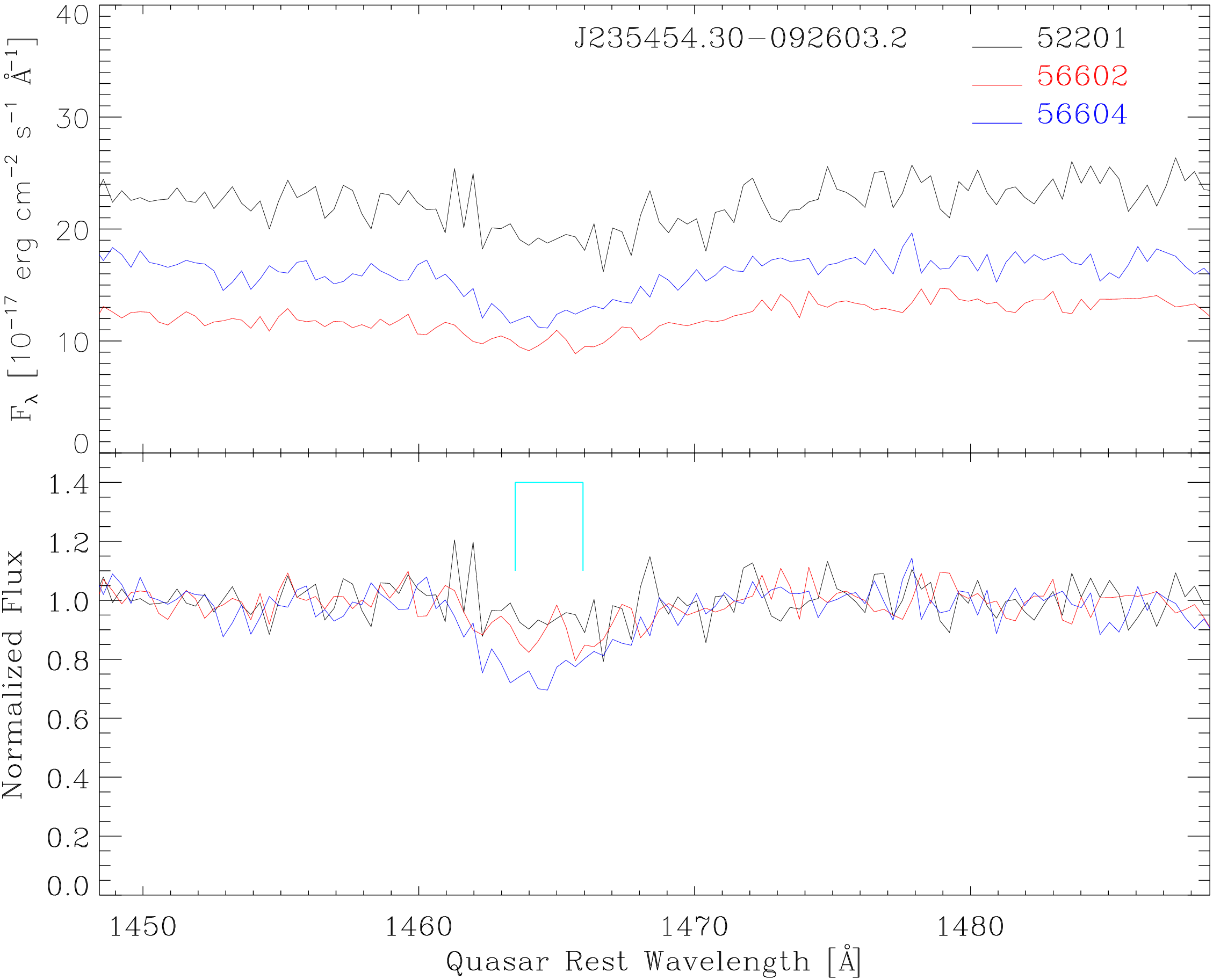}
\caption{The quasar spectra with at least one \CIV\ NAL meeting $\mid\Delta W_r^{\lambda1548}\mid\ge3\sigma_{\rm \Delta W_r^{\lambda1548}}$. In each figure, the upper panel is the SDSS spectra, and the lower panel is the SDSS spectra normalized by the pseudo-continuum fits. Different color lines indicate the spectra obtained at different times. The significantly variable \CIV\ NALs are marked by cyan lines. The values in the top right corners are the spectra MJD.}
\label{fig:variableabs}
\end{figure*}

\end{document}